\documentclass[twocolumn]{aastex63}
\usepackage{graphicx}
\usepackage{amsmath}
\usepackage{amssymb}
\usepackage{cases}
\usepackage{mathrsfs}
\usepackage[flushleft]{threeparttable}
\usepackage{amssymb}
\usepackage{pifont}
\usepackage{amsmath}
\usepackage{epstopdf}
\usepackage{xcolor}
\usepackage[all]{hypcap}
\usepackage{mwe,tikz}
\usepackage{bm}
\usepackage{tabularx}

\usepackage{xspace}
\usepackage{rotating}

\newcommand{\e}{\text{e}}

\newcommand{\p}{\partial}

\newcommand{\sw}[1]{\texttt{#1}}

\received{2020 June 23}
\revised{2020 July 10}
\accepted{2020 July 16}
\submitjournal{ApJ}

\shorttitle{Implications of GRB 200415A}
\shortauthors{Yang et al.}

\begin{document}

\title{GRB 200415A: A Short Gamma-Ray Burst from a Magnetar Giant Flare?}
\correspondingauthor{Bin-Bin Zhang and Vikas Chand}
\email{bbzhang@nju.edu.cn and vikasK2@nju.edu.cn}

\author[0000-0002-5485-5042]{Jun Yang}
\affiliation{School of Astronomy and Space Science, Nanjing
University, Nanjing 210093, China}
\affiliation{Key Laboratory of Modern Astronomy and Astrophysics (Nanjing University), Ministry of Education, China}

\author[0000-0002-7876-7362]{Vikas Chand}
\affiliation{School of Astronomy and Space Science, Nanjing
University, Nanjing 210093, China}
\affiliation{Key Laboratory of Modern Astronomy and Astrophysics (Nanjing University), Ministry of Education, China}

\author[0000-0003-4111-5958]{Bin-Bin Zhang}
\affiliation{School of Astronomy and Space Science, Nanjing
University, Nanjing 210093, China}
\affiliation{Key Laboratory of Modern Astronomy and Astrophysics (Nanjing University), Ministry of Education, China}
\affiliation{Department of Physics and Astronomy, University of Nevada Las Vegas, NV 89154, USA}

\author[0000-0003-0691-6688]{Yu-Han Yang}
\affiliation{School of Astronomy and Space Science, Nanjing
University, Nanjing 210093, China}
\affiliation{Key Laboratory of Modern Astronomy and Astrophysics (Nanjing University), Ministry of Education, China}

\author{Jin-Hang Zou}
\affiliation{College of Physics, Hebei Normal University, Shijiazhuang 050024, China}

\author[0000-0002-7555-0790]{Yi-Si Yang}
\affiliation{School of Astronomy and Space Science, Nanjing
University, Nanjing 210093, China}
\affiliation{Key Laboratory of Modern Astronomy and Astrophysics (Nanjing University), Ministry of Education, China}

\author{Xiao-Hong Zhao}
\affiliation{Yunnan Observatories, Chinese Academy of Sciences, Kunming, China}
\affiliation{Center for Astronomical Mega-Science, Chinese Academy of Sciences, Beijing, China}
\affiliation{Key Laboratory for the Structure and Evolution of Celestial Objects, Chinese Academy of Sciences, Kunming, China}

\author{Lang Shao}
\affiliation{College of Physics, Hebei Normal University, Shijiazhuang 050024, China}

\author{Shao-Lin Xiong}
\affiliation{Key Laboratory of Particle Astrophysics, Institute of High Energy Physics, Chinese Academy of Sciences, 19B Yuquan Road, Beijing 100049, People’s Republic of China}

\author{Qi Luo}
\affiliation{Key Laboratory of Particle Astrophysics, Institute of High Energy Physics, Chinese Academy of Sciences, 19B Yuquan Road, Beijing 100049, People’s Republic of China}
\affiliation{University of Chinese Academy of Sciences, Chinese Academy of Sciences, Beijing 100049, China}

\author{Xiao-Bo Li}
\affiliation{Key Laboratory of Particle Astrophysics, Institute of High Energy Physics, Chinese Academy of Sciences, 19B Yuquan Road, Beijing 100049, People’s Republic of China}

\author{Shuo Xiao}
\affiliation{Key Laboratory of Particle Astrophysics, Institute of High Energy Physics, Chinese Academy of Sciences, 19B Yuquan Road, Beijing 100049, People’s Republic of China}

\author{Cheng-Kui Li}
\affiliation{Key Laboratory of Particle Astrophysics, Institute of High Energy Physics, Chinese Academy of Sciences, 19B Yuquan Road, Beijing 100049, People’s Republic of China}

\author{Cong-Zhan Liu}
\affiliation{Key Laboratory of Particle Astrophysics, Institute of High Energy Physics, Chinese Academy of Sciences, 19B Yuquan Road, Beijing 100049, People’s Republic of China}

\author[0000-0003-3383-1591]{Jagdish C. Joshi}
\affiliation{School of Astronomy and Space Science, Nanjing
University, Nanjing 210093, China}
\affiliation{Key Laboratory of Modern Astronomy and Astrophysics (Nanjing University), Ministry of Education, China}

\author[0000-0002-4394-4138]{Vidushi Sharma}
\affiliation{Inter University Centre for Astronomy and Astrophysics, Pune, India}

\author{Manoneeta Chakraborty}
\affiliation{DAASE, Indian Institute of Technology Indore, Khandwa Road, Simrol, Indore 453552, India}

\author[0000-0001-5931-2381]{Ye Li}
\affiliation{Kavli Institute for Astronomy and Astrophysics, Peking University, Beijing 100871, China}
\affiliation{Purple Mountain Observatory, Chinese Academy of Sciences, Nanjing 210008, China}

\author[0000-0002-9725-2524]{Bing Zhang}
\affiliation{Department of Physics and Astronomy, University of Nevada Las Vegas, NV 89154, USA}

\begin{abstract}

The giant flares of soft gamma-ray repeaters (SGRs) have long been proposed to contribute to at least a subsample of the observed short gamma-ray bursts (GRBs). In this paper, we perform a comprehensive analysis of the high-energy data of the recent bright short GRB 200415A, which was located close to the Sculptor galaxy. Our results suggest that a magnetar giant flare provides the most natural explanation for most observational properties of GRB 200415A, including its location, temporal and spectral features, energy, statistical correlations, and high-energy emissions. On the other hand, the compact star merger GRB model is found to have difficulty reproducing such an event in a nearby distance. Future detections and follow-up observations of similar events are essential to firmly establish the connection between SGR giant flares and a subsample of nearby short GRBs.

\end{abstract}


\section{Introduction} \label{sec:intro}

With the joint detection of the gravitational-wave event GW 170817 and the short gamma-ray burst (sGRB) GRB 170817A, mounting evidence has been established to support that at least some short GRBs can originate from the merger of two compact stars in, e.g., binary neutron star or black hole-neutron star systems. Nevertheless, such an association still allows the alternative model that giant flares (GFs) from magnetars (manifested as soft gamma-ray repeaters (SGRs)) in nearby galaxies can also produce sGRB-like events, even though only GRB 051103 \citep{Ofek:2006ApJ.652.507O, Frederiks:2007AstL33} and GRB 070201 \citep{Mazets:2008ApJ, Ofek:2008ApJ} have been proposed as candidates for GF-sGRBs in the past.

The discovery of SGRs in 1979 \citep[e.g.,][]{Mazets:1979SvAL, Mazets:1979Nature} suggested that some short hard X-ray and soft gamma-ray bursts with highly nonuniform periods came from the same source. Observations have revealed that different kinds of bursts can be detected from SGRs, namely recurrent bursts and GFs. Recurrent bursts can be relatively common during the reactivation periods after the long-term quiescence of SGRs. The spectral temperature of about 25--35 keV obtained from the optically thin thermal bremsstrahlung (OTTB) model often characterizes the recurrent bursts in the 20--200 keV range \citep{Mazets:1982Ap&SS, Golenetskii:1984Nature, Laros:1986Nature, Woods:1999ApJ}. In previous time-integrated spectral analyses, the spectra of the most recurrent bursts can be described by thermal models (e.g. OTTB, blackbody(BB), or the sum of two BBs), nonthermal models (e.g. power law or cutoff power law), or some more complex models \citep{van-der-Horst:2012ApJ, Kirmizibayrak:2017ApJS}. A GF is another type of SGR activity, but rarer and more powerful. There are only three confirmed GFs discovered since 1979 from three SGRs: SGR 0526-66 \citep{Mazets:1979Nature, Mazets:1982Ap&SS}, SGR 1900+14 \citep{Cline:1998IAUC, Hurley:1999Nature, Kouveliotou:1999ApJ, Mazets:1999AstL}, and SGR 1806-20 \citep{Hurley:2005Nature, Mereghetti:2005ApJ.624L.105M, Palmer:2005Nature, Frederiks:2007AstL}. The initial pulse is the most intense component of the GF, which is usually characterized by an abrupt rise and a quasi-exponential decay in the light curve and hard spectra that rapidly evolve with time \citep{Hurley:1999Nature, Mazets:1999AstL, Frederiks:2007AstL}. The subsequent pulsating tail represents some quasi-periodic oscillation (QPO) behaviors \citep{Israel_2005, Strohmayer_2005}. Its spectral temperature is similar to those of recurrent bursts. The intensities of pulsating bursts often show an exponential decay \citep{Mazets:1999AstL}. In addition, a peculiar burst detected in SGR 1627-41 behaved very similarly to other GFs when its intensity, spectral properties, and energy are taken into account, even though it ended with a single pulse and exhibited no steep rise in the light curve \citep{Mazets:1999ApJ, Woods:1999ApJ}.

Some SGRs are coincident with young ($\sim 10^4 \rm\,yr$) supernova remnants \citep{Cline:1982ApJ, Kulkarni:1993Natur} or located in regions with a high star formation rate. Observations show that SGRs are detected as a continuous X-ray radiation source \citep{Murakami:1994Natur} with periodic light curves during its quiescent phase. The magnetar model \citep{Duncan:1992ApJ.392L, Thompson:1995MNRAS} has long been proposed to generate SGRs. In the context of the magnetar model, the internal heating caused by the decaying magnetic field \citep[e.g.,][]{Thompson:1996ApJ.473.322T} or the currents in the twisted global magnetosphere \citep{Thompson:2002ApJ} sustained by the twisting motion \citep{Thompson:2000ApJ.543.340T} of the magnetar crust have been suggested to explain the X-ray emission from SGRs. The SGR bursts can be energized during starquakes that result from fractures of the neutron star crust when the crust undergoes strong magnetic stresses \citep{Thompson:1995MNRAS}. In this scenario, the neutron star is characterized by a dipolar magnetic field of $\sim10^{14}$--$10^{15}$ G \citep{Duncan:1992ApJ.392L}. \cite{Thompson:2002ApJ} proposed the presence of a twisted magnetosphere, which will increase the spin-down. Large-scale magnetic reconnection can take place in a twisted magnetosphere \citep{Parfrey_2013} to heat the magnetic corona \citep{Lyutikov:2003MNRAS.346.540L}. The helical distortion or interchange instability of the interior magnetic field ruptures the magnetar crust, and then the large-scale twisting exterior magnetic field in the global magnetosphere may lead to the dissipation of enormous magnetic energy to produce a GF \citep{Thompson_2001}.

The typical energy of a magnetar GF is $\sim 10^{44}$--$10^{46}$ erg \citep{Mazets:1999AstL, Hurley:2005Nature}. At a nearby galaxy with a distance of a few Mpc, such energy corresponds to a fluence level of $10^{-8}$--$10^{-5}$ $\rm erg\,cm^{-2}$, which is roughly consistent with the observed fluence of short GRBs \citep{2016ApJS..223...28N}. Additionally, the hard spikes of SGR giant flares are also similar to short GRBs in terms of temporal and spectral properties \citep[e.g.,][]{Mazets:1982Ap&SS}. The distant extragalactic GFs without detectable tails are suggested to be a subset of short GRBs with a nonnegligible fraction \citep{Duncan:2001AIPC.586.495D, Hurley:2005Nature, Lazzati:2005MNRAS}. For these reasons, great interest has been shown in the possibility of an SGR GF origin once any short GRBs are detected at a location consistent with a nearby galaxy.

Recently, an extremely bright short-duration GRB, GRB 200415A, triggered the Fermi Gamma-ray Burst Monitor \citep[GBM;][]{Meegan:2009ApJ} at 08:48:05.564 on 2020 April 15 UTC \citep[hereafter $T_0$;][]{Bissaldi:2020GCN}. This GRB was also detected by the High energy X-ray Telescope loaded on the Hard X-ray Modulation Telescope \citep[HXMT-HE;][]{Li:2007NuPhS, Zhang:2020SCPMA}, although the detectors were saturated early. Combining Konus-Wind and Swift-BAT data, the Inter-Planetary Network (IPN) located the burst source at $\alpha=00^{\rm h}47^{\rm m}30^{\rm s}$ and $\delta=-25^{\circ}11^{\prime}37^{\prime\prime}$ (J2000), with an error box area of 274 arcmin$^2$ \citep{Svinkin:2020GCN}. This location lies close to the Sculptor galaxy (aka NGC 253), which has a distance of about $3.5$ Mpc. The Sculptor galaxy is the seventh-brightest galaxy (besides the Milky Way) in the sky, with a half-light radius of $4^{\prime}$ \citep{2003AJ....125..525J} and a distance $5^{\prime}.7$ to the center of the IPN position of GRB 200415A. Following \cite{bloom2002}, the probability of a random 274 arcmin$^2$ region (3$\sigma$) on the sky falling so close to a galaxy with similar magnitude is $P_{\rm ch}=1.3\times10^{-5}$. Such an association likely points to an SGR GF origin. Unlike previous GF-sGRB candidates that only have relatively limited information, we can investigate this event in unprecedented detail thanks to the high temporal and spectral resolution of the Fermi/GBM data in the hope of confirming that it was indeed an SGR GF but misclassified as a typical short GRB. To do so, we perform a comprehensive analysis of Fermi (including GBM and LAT) data on this burst, as shown in \S \ref{sec:fermi_data_ana}. In \S \ref{sec:physical_origin}, we discuss its classification. This is followed by a summary and implications in \S \ref{sec:summary}.

\section{Fermi Data Analysis} \label{sec:fermi_data_ana}

We retrieved the burst time-tagged event data set that covers the time range of GRB 200415A from the Fermi-GBM public data archive\footnote{\url{https://heasarc.gsfc.nasa.gov/FTP/fermi/data/gbm/daily/}}. Data reduction and analysis follow the procedures discussed in \cite{Zhang:2011ApJ730, Zhang:2016ApJ816, Zhang:2018NatAs}. The temporal and spectral properties of GRB 200415A are listed below.

\subsection{Light Curves and Timescale} \label{sec:lighcurves}

\begin{figure}
 \label{fig:wave}
 \includegraphics[width=0.47\textwidth]{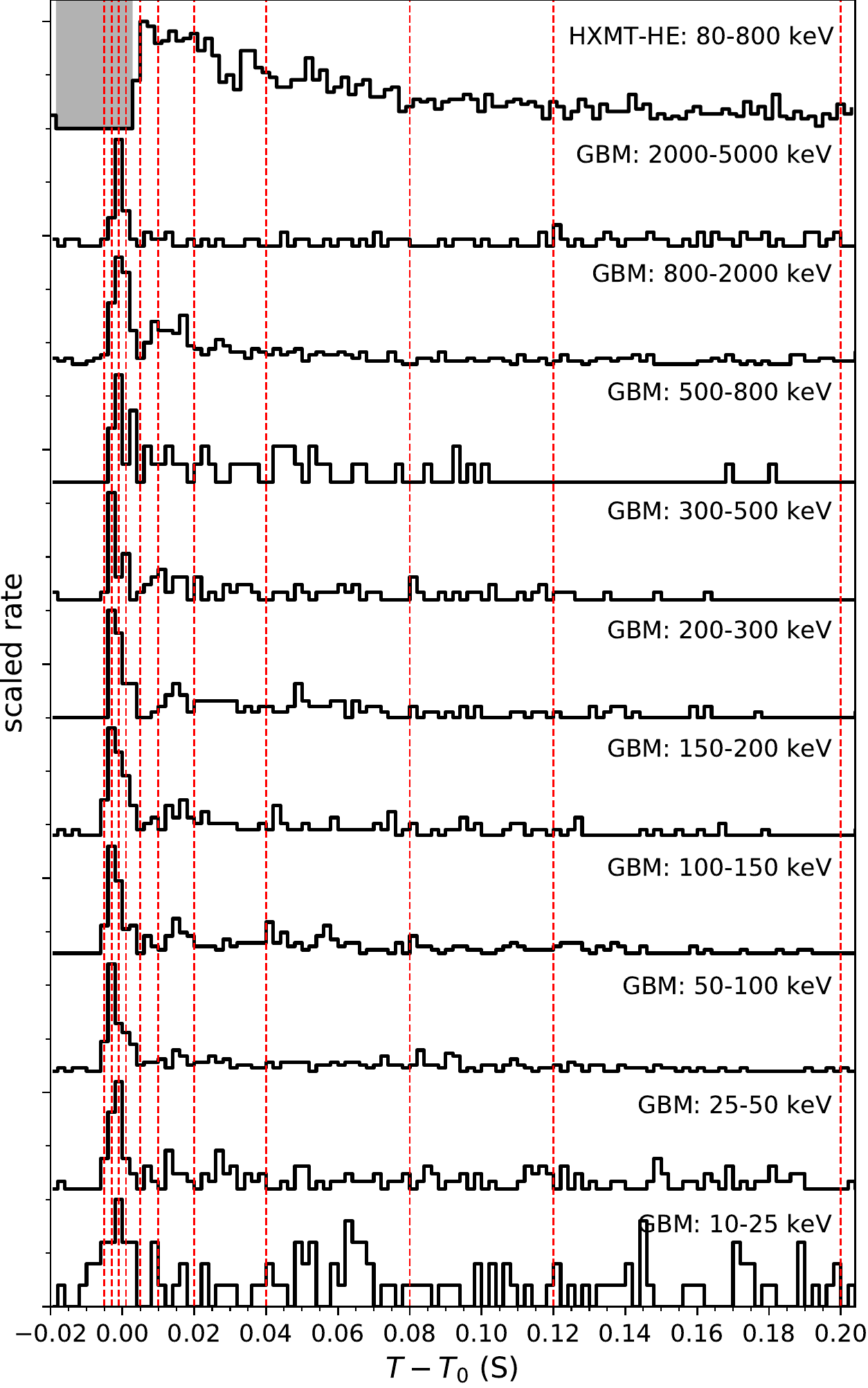}
 \caption{Multiwavelength light curves obtained by combining the data from the Fermi/GBM NaI detector n1, Fermi/GBM BGO detector b0, and HXMT-HE detectors. The red vertical lines mark several time slices (as listed in Table \ref{tab:spe_para}) for spectral analyses. The gray shaded area corresponds to the early saturation time (about 21 ms) of the HXMT-HE detectors.}
\end{figure}

\begin{figure}
 \centering
 \includegraphics[width=0.47\textwidth]{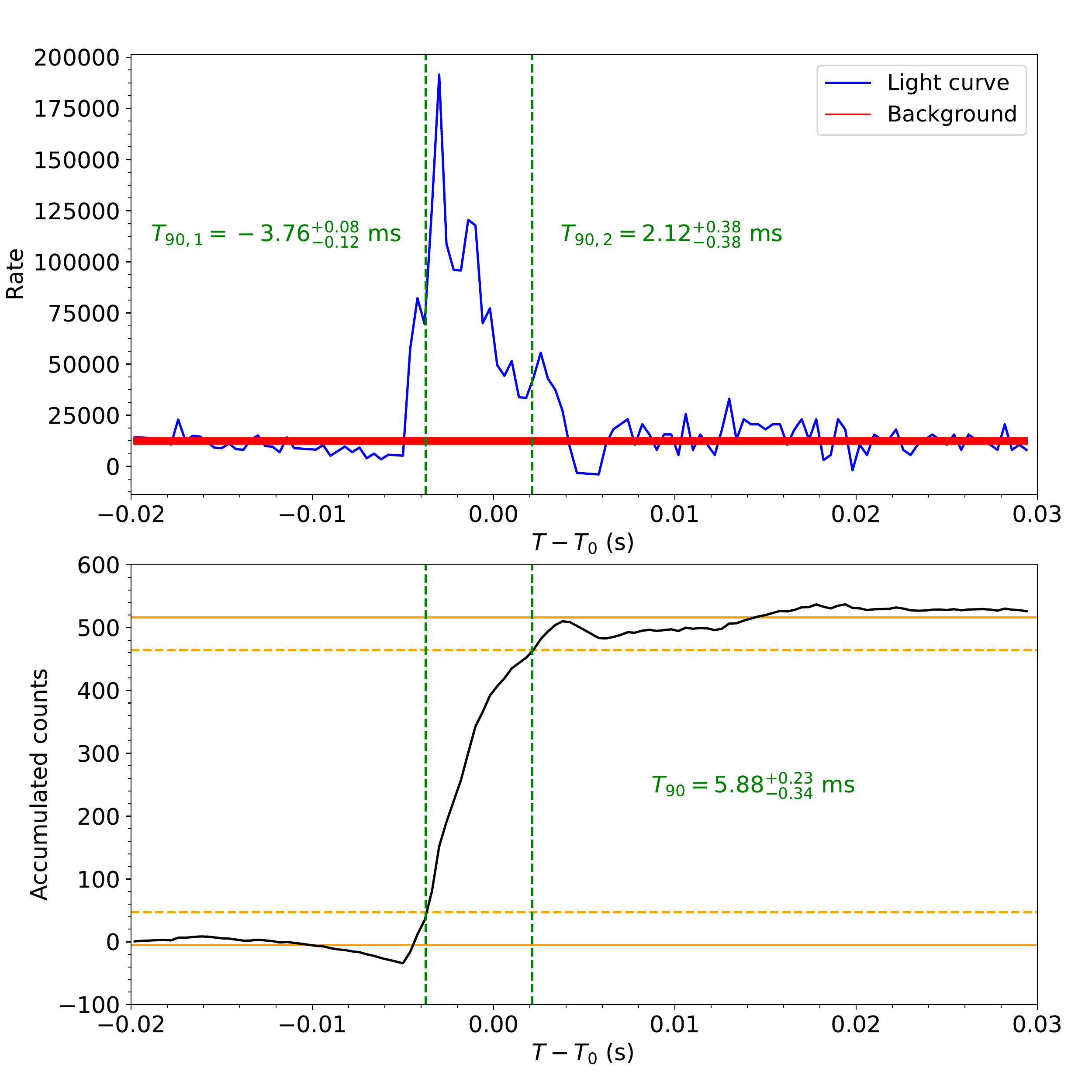} 
 \caption{The $T_{90}$ calculation. In the upper panel, the blue line shows the light curve obtained by combining the GBM-n1 and GBM-n3 data. The level of background is represented by the red line. In the lower panel, the accumulated counts are shown with the black line. The orange horizontal dashed (solid) lines are drawn at 5\% (0\%) and 95\% (100\%) of the total accumulated counts. In both panels, the $T_{90}$ interval is marked by the green vertical dashed lines.}
 \label{fig:T_90}
\end{figure}

\begin{figure}[t]
\centering
\includegraphics[angle=0,width=0.47\textwidth]{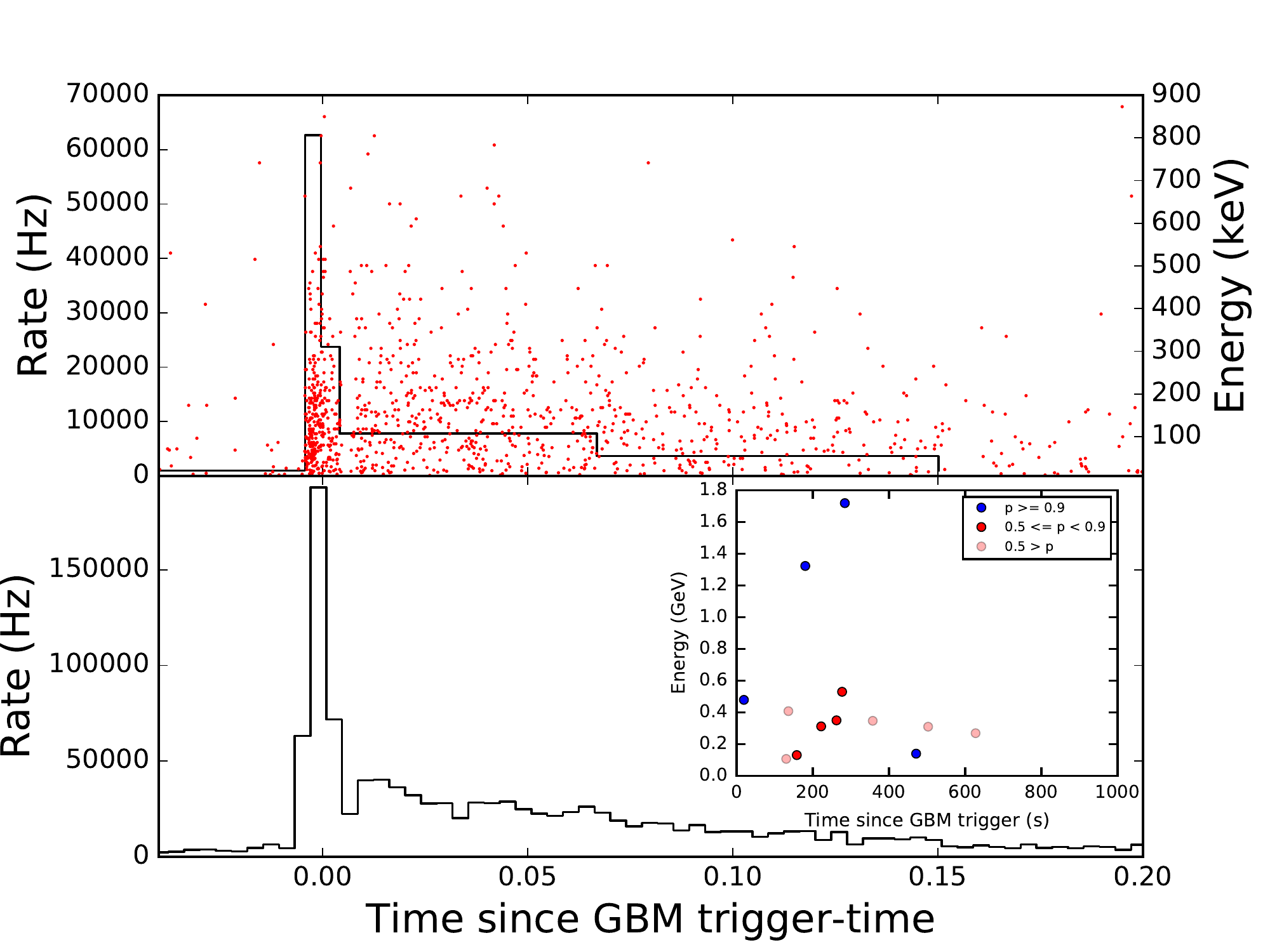}
\caption{Light curves of GRB 200415A. Upper panel: Bayesian block light curve for GBM-n1 and energy scatters in the 8--900 keV range. Lower panel: combined light curve from the NaI detectors (n0, n1, n3, and n5) with the minimum Bayesian block bin size (3.8 ms). Inset: LAT-HE observations in 0--1000 s.}
\label{fig:bb_lcs}
\end{figure}

The multiwavelength light curves of this burst are shown in Figure \ref{fig:wave} and derived from the photons collected by sodium iodide (NaI) detector n1 and bismuth germanium oxide (BGO) detector b0. The background is modeled via applying the ``baseline" method \citep{Zhang:2018NatAs} to a wide time interval around the signal and subtracted in GBM light curves. The light curve obtained from HXMT-HE within 80--800 keV is added to the topmost panel in Figure \ref{fig:wave}. The correction of the aberration of light effect, which is 15.04 ms in terms of photon arrival time difference when they reached the GBM and HXMT-HE detectors, has been applied on the HXMT data. The count rates in the HXMT-HE light curve have also been corrected for the dead time and saturation \citep{Xiao:2020JHEAp}. The early 21 ms saturation time of HXMT-HE is shown with a gray shaded area in Figure \ref{fig:wave}. The light curves show a sharp peak at $\sim T_0$ with a duration of $T_{90}=5.88_{-0.34}^{+0.23}$ ms (see Figure \ref{fig:T_90}), which consists of an abrupt rise ($\sim2$ ms) and a steep decay ($\sim8$ ms). The sharp peak is followed by a soft tail (also see Figure \ref{fig:bb_lcs}) that extends to $\sim$ $T_0+0.2$ s in lower energies.

\begin{figure}
\centering
\includegraphics[angle=0,width=0.48\textwidth]{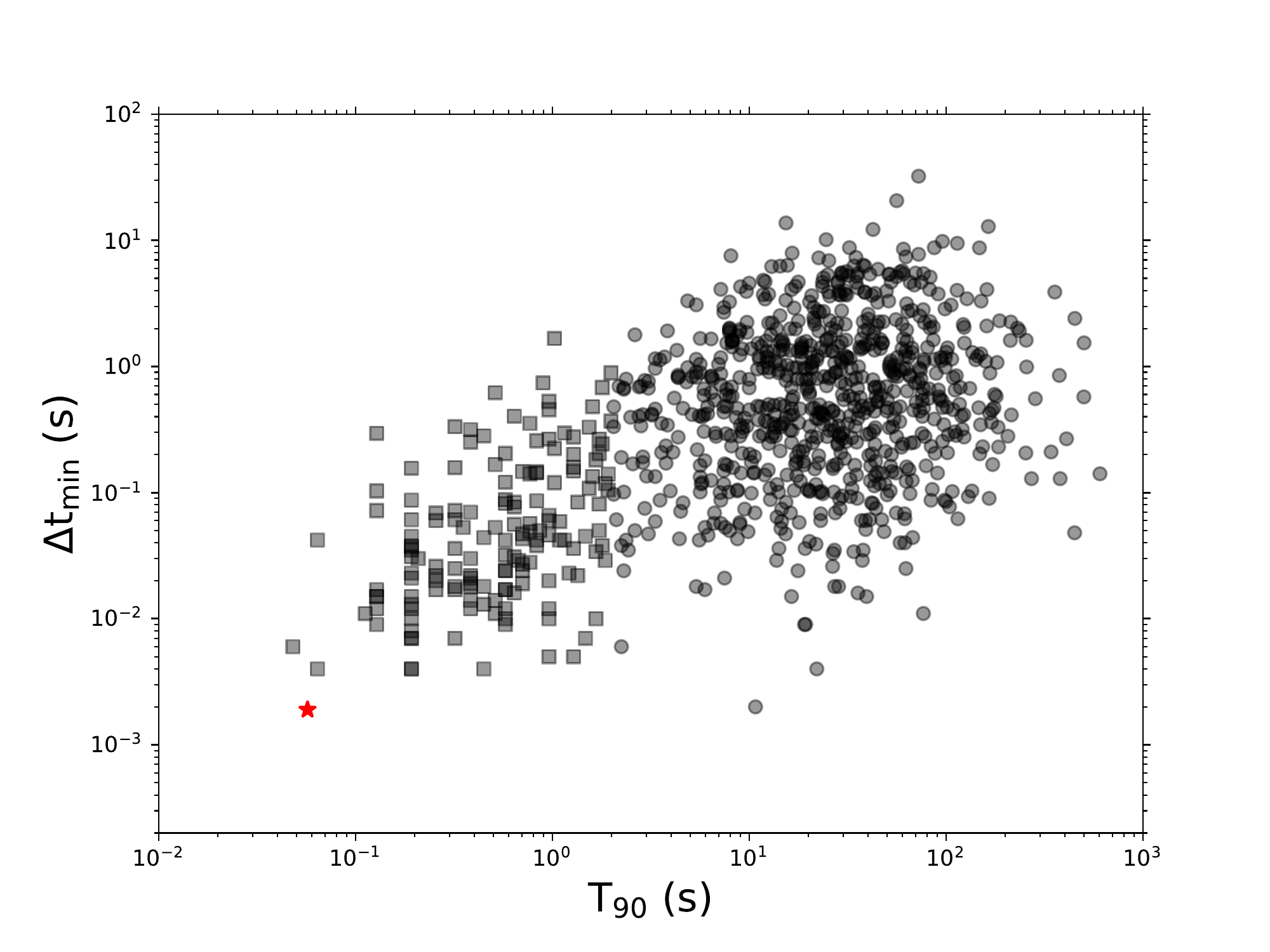}
\caption{The $T_{90}$--$\Delta t_{\rm min}$ diagram. The sample of short (square) and long (circle) GRBs is obtained from \cite{Golkhou:2015ApJ}, and GRB 200415A is highlighted by a red star.} 
\label{fig:mvts_T90}
\end{figure}

To determine the minimal time variability ($\Delta t_{\rm min}$), we employ the Bayesian block \citep{scargle:2013} method on the photon events between 8 and 900 keV. The resulting blocks, as shown in the upper panel of Figure \ref{fig:bb_lcs}, can track the statistically significant changes in the light curve. The minimum bin size of the obtained blocks is 3.8 ms. We regard half of the minimum bin size as the minimal time variability of this burst \citep[e.g.,][]{Vianello:2018ApJ.864.163V}. We note that it is one of the smallest $\Delta t_{\rm min}$ among all GRBs in Figure \ref{fig:mvts_T90} where we plot a sample of short and long GRBs in the $T_{90}$--$\Delta t_{\rm min}$ plane obtained from \cite{Golkhou:2015ApJ}.

The combined light curve obtained from the NaI detectors (n0, n1, n3, and n5), binned at 3.8 ms, is shown in the lower panel of Figure \ref{fig:bb_lcs}. It is also clear that the high-energy emission starts and ends at $\sim T_0-0.005$ s and $\sim T_0+0.2$ s, respectively, which consists of a first spike and a weak tail. Such a time range will be used for the spectral lag calculations in \S\ref{sec:lag} and the spectral analyses in \S\ref{sec:spectal_ana}.

\subsection{Spectral Lags} \label{sec:lag}

\begin{figure}
 \label{fig:lag}
 \centering
 \includegraphics[width=0.47\textwidth]{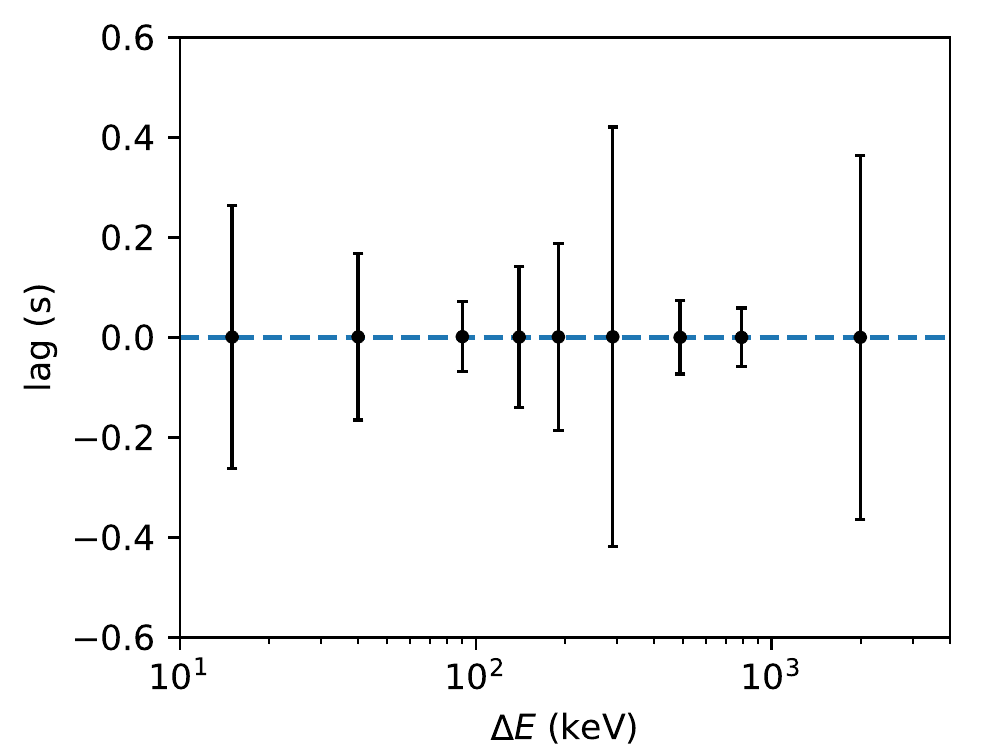}
 \caption{Time lags between any higher energy band and the lowest energy band in the GBM multiwavelength light curves. Here $\Delta E$ is the difference of the lower limits of two energy bands. The zero lag is shown with a dashed blue line.}
\end{figure}

Systematic time lags among light curves in different energy bands have been observed in large GRB samples \citep[e.g.,][]{Yi:2006, Ukwatta_2010, Shao:2017ApJ}. A curvature effect has been suggested as a cause of spectral lags in GRB prompt emission \citep[e.g.,][]{Ioka:2001ApJ, Norris:2002ApJ, Shenoy:2013ApJ}. However, it has recently been found that the observed spectral lags can not be explained by the curvature effect alone, and they are also related to a curved photon spectrum, decreasing magnetic field, and large emission region undergoing a rapid bulk acceleration \citep{Uhm:2016ApJ.825.97U}. The existence of spectral lags supports that the prompt emission in GRBs is generated via a Poynting flux-dominated jet that abruptly dissipates magnetic energy at a large distance from the engine \citep{Uhm:2016ApJ.825.97U}. Compared to long GRBs, short GRBs show negligible lags \citep{Bernardini:2015MNRAS}. For GRB 200415A, we select the time interval from $T_0-0.005$ s to $T_0+0.20$ s; then, we follow \cite{Zhang_2012} to use the cross-correlation function \citep{Norris:2000ApJ.534.248N, Ukwatta_2010} to measure the time lags between any higher energy band and the lowest energy band in the GBM multiwavelength light curves, as shown in Figure \ref{fig:wave}. The uncertainties of time lags depend on the selected time interval and are estimated by Monte Carlo simulations \citep[e.g.,][]{Ukwatta_2010, Zhang_2012}. Interestingly, our results show that all of the lag values are very tiny ($\lesssim$ 1 ms) and consistent with zero lag (Figure \ref{fig:lag}) when considering their uncertainties. Such a result is similar to previous studies \citep[see, e.g.,][]{Zhang:2009ApJ.703.1696Z}, where short GRBs were indicated with tiny lags. The tiny-lag result of GRB 200415A suggests that this event was likely not from a GRB jet but rather from a one-time fireball-like emission region. The spectral analyses in the next subsection confirm this suggestion.

\subsection{Spectral Analysis}\label{sec:spectal_ana}

We perform both time-integrated and time-dependent spectral analyses between $T_0-0.005$ s and $T_0+0.20$ s. This time interval is divided into thirteen slices (see Table \ref{tab:spe_para} or Figure \ref{fig:wave}) according to the brightness and the count statistical significance for spectral fitting \citep{Zhang:2018NatAs}, in which three slices (covering the first spike, weak tail, and total emission) are used for time-integrated spectral analyses, and others are used for time-dependent spectral analyses. Within each slice, we extract the spectra of GRB 200415A from two NaI detectors (n1, n3) that are within a 60$^{\circ}$ angle with respect to the burst and BGO detector b0. Corresponding background spectra are acquired by applying the ``baseline" method \citep{Zhang:2018NatAs} to the time interval from $T_0-20$ s to $T_0+20$ s for each energy channel. The response matrices of the detectors are generated using the response generator provided by the \sw{GBM Response Generator}\footnote{\url{https://fermi.gsfc.nasa.gov/ssc/data/analysis/rmfit/gbmrsp-2.0.10.tar.bz2}}. Then we use \sw{McSpecfit} \citep{Zhang:2018NatAs} to perform the spectral fitting. Several spectral models, including single power law (PL), cutoff power law (CPL), band function (Band), BB, and their combinations, are adopted. Additionally, a multicolor BB (mBB) model is also used in this study. Assuming that the distribution of the thermal luminosity with temperature follows the PL of index $m$, the mBB model can be formulated as \citep{Hou:2018ApJ.866.13H}
\begin{equation}
 N(E)=\frac{8.0525(m+1)K}{\Big[\big(\frac{T_{\rm max}}{T_{\rm min}}\big)^{m+1}-1\Big]}\Big(\frac{kT_{\rm min}}{\rm keV}\Big)^{-2}I(E),\label{N(E)}
\end{equation}
where
\begin{equation}
 I(E)=\Big(\frac{E}{kT_{\rm min}}\Big)^{m-1}\int_{\frac{E}{kT_{\rm max}}}^{\frac{E}{kT_{\rm min}}}\frac{x^{2-m}}{e^x-1}dx,\label{I(E)}
\end{equation}
$x=E/kT$, and $K=L_{39}/D_{\rm L,10\rm kpc}^2$, with the luminosity $L$ in units of $10^{39}$ $\rm erg\,s^{-1}$ and the luminosity distance $D_{\rm L}$ in units of 10 kpc.

Based on the Bayesian information criteria (BIC), we determine the best model for each time interval. The best-fit parameters of the CPL, BB, and mBB models for all time slices are listed in Table \ref{tab:spe_para}.

\begin{figure*}
\begin{tabular}{lll}
\includegraphics[angle=0,width=0.3\textwidth]{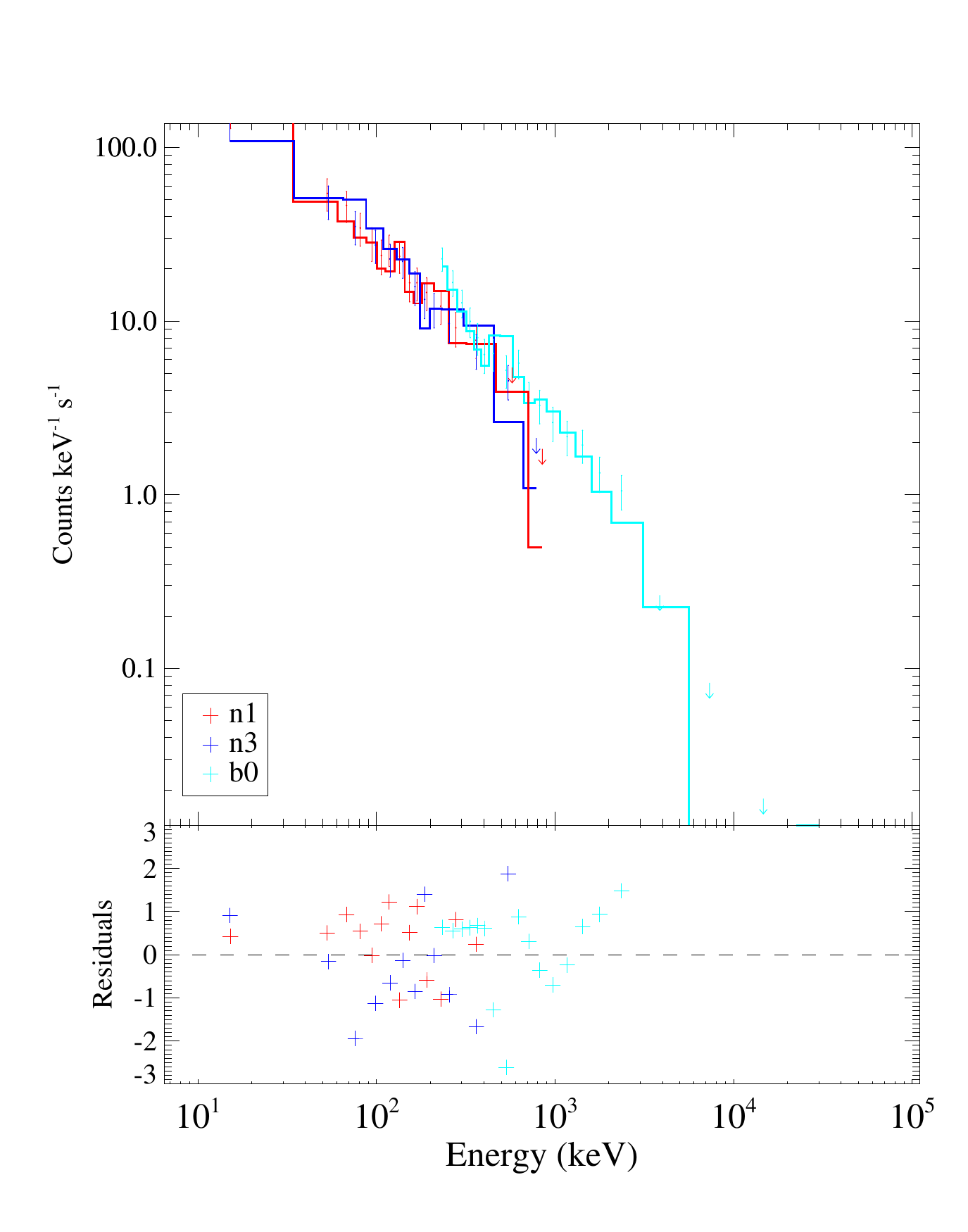} &
\includegraphics[angle=0,width=0.3\textwidth]{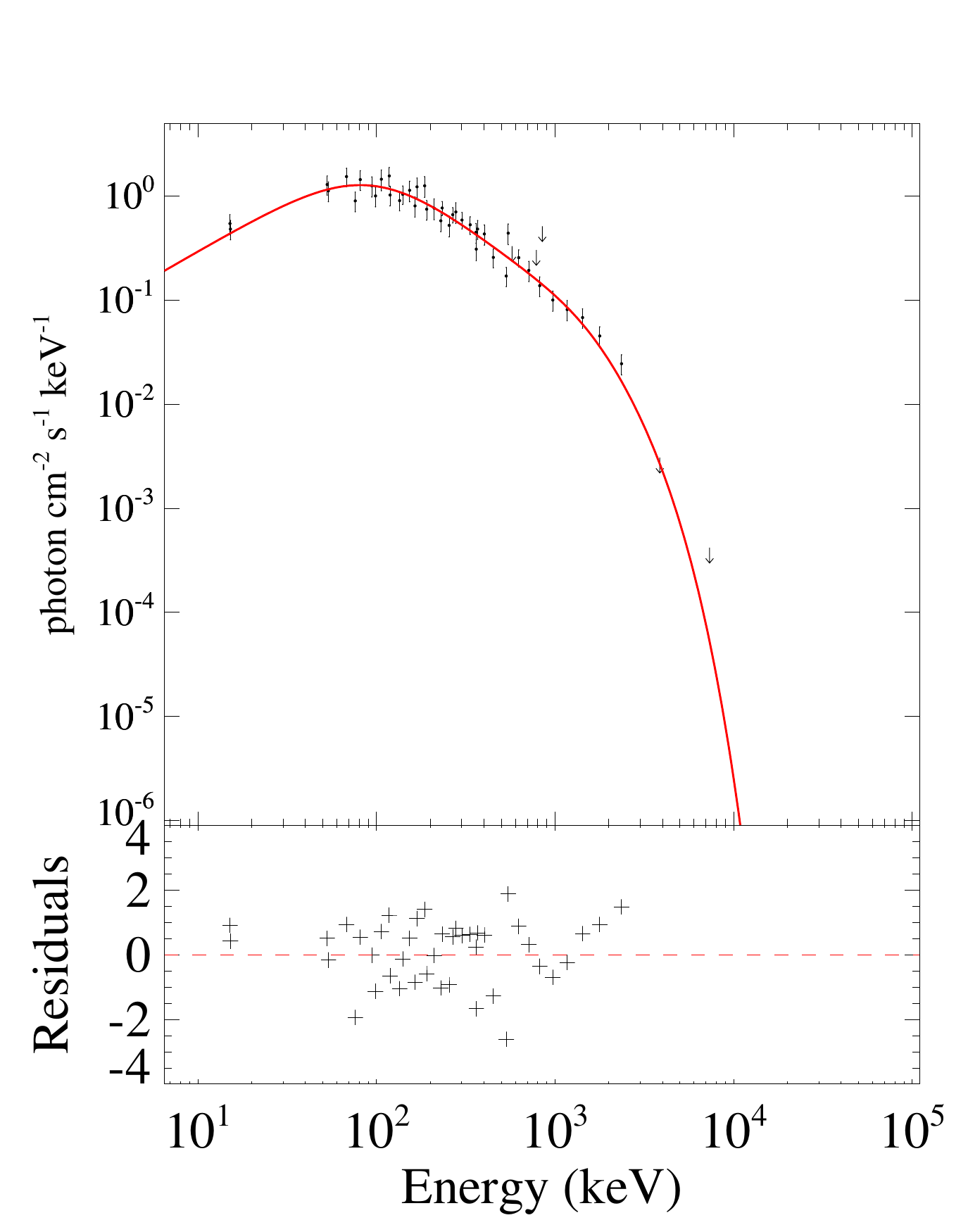} &
\includegraphics[angle=0,width=0.3\textwidth]{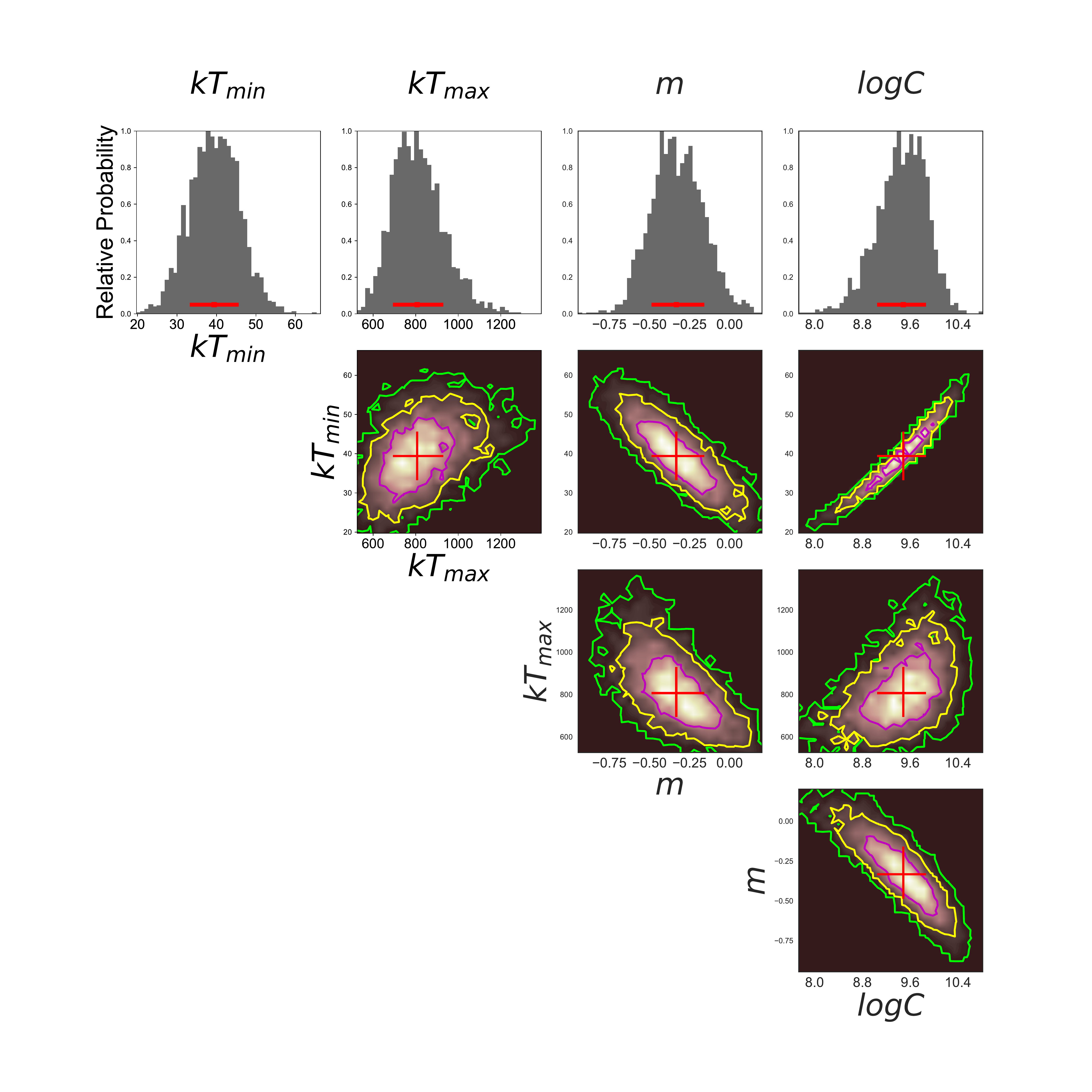}
\end{tabular}
\caption{Spectral fitting of the mBB model for the time interval from $T_0-0.005$ s to $T_0+0.005$ s. Left: observed photon count spectra and the best-fit model. Middle: deconvolved photon spectrum. Right: corner diagram for the parameters of the mBB model. Histograms show the 1D probability distributions for the parameters. Contours illustrate the likelihood 2D map. Red plus signs mark the best-fitting values. All error bars in these panels represent the 1$\sigma$ credible intervals.}
\label{fig:mbb_corner}
\end{figure*}

\begin{figure*}
\begin{tabular}{lll}
\includegraphics[angle=0,width=0.3\textwidth]{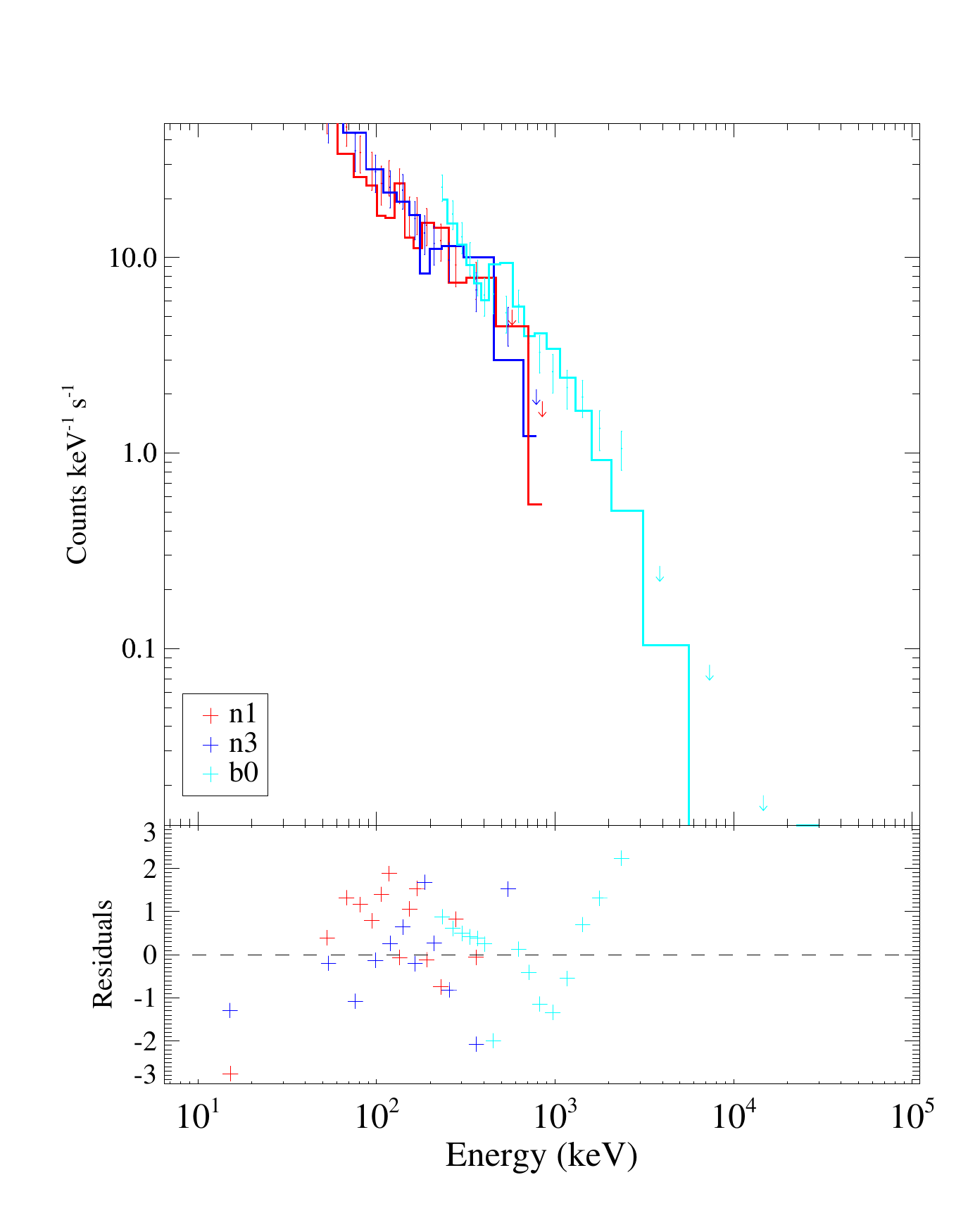}&
\includegraphics[angle=0,width=0.3\textwidth]{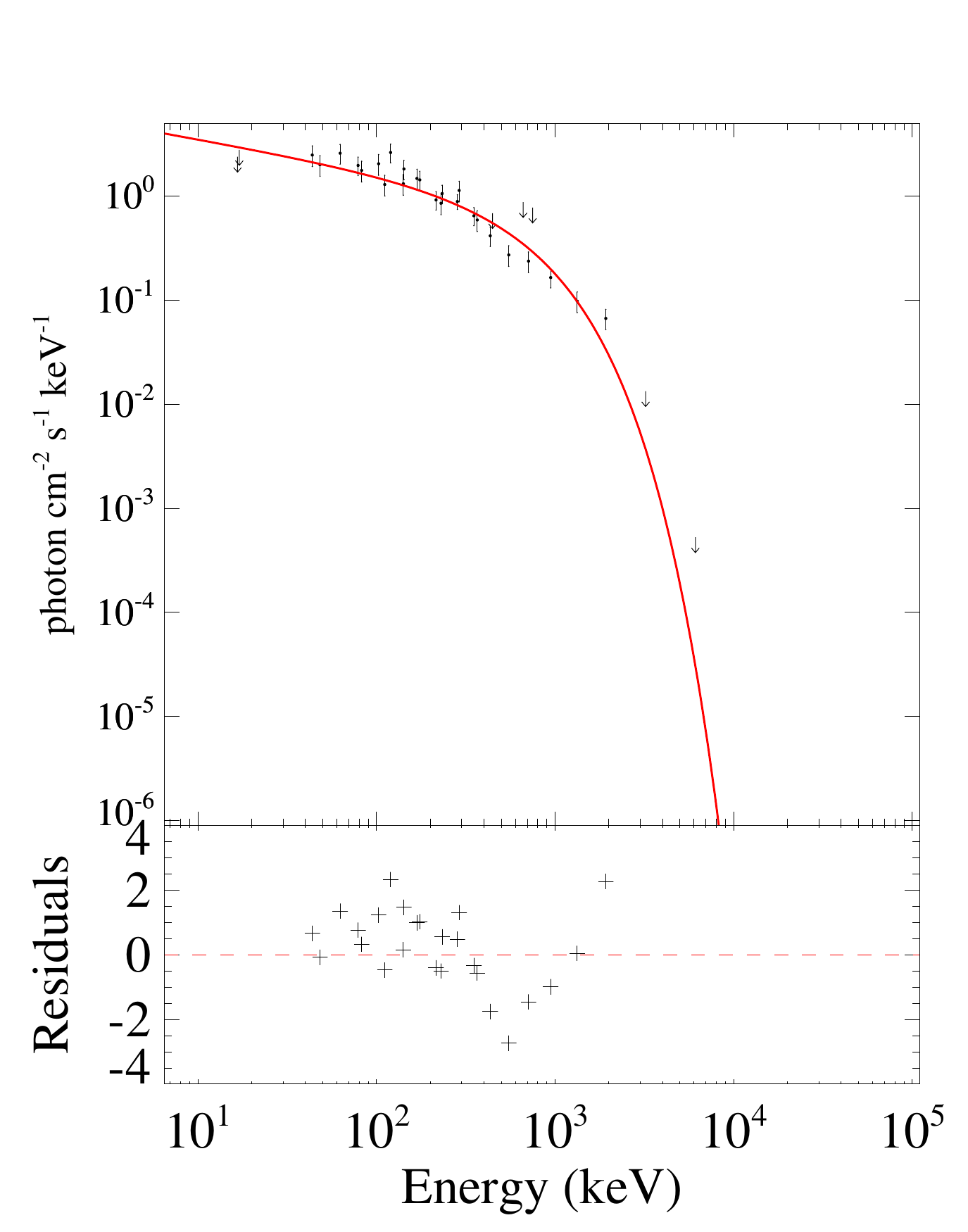}&
\includegraphics[angle=0,width=0.3\textwidth]{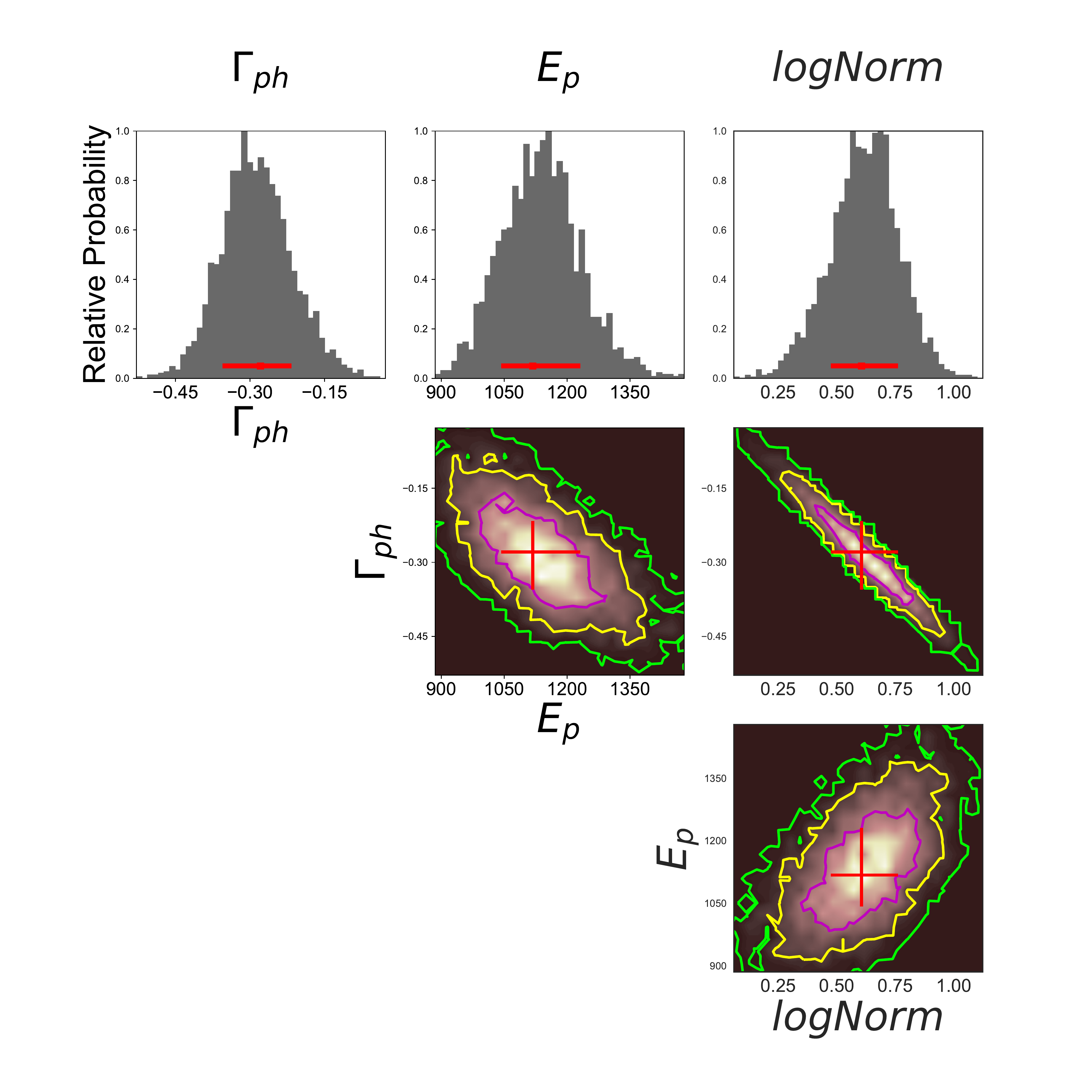}
\end{tabular}
\caption{Same as Figure \ref{fig:mbb_corner} but fitted by the CPL model.}
\label{fig:cpl_corner}
\end{figure*}

The time-integrated spectrum measured between $T_0-0.005$ s and $T_0+0.005$ s, covering the first spike, can be best fitted by the mBB model with $kT_{\rm min}=39.40_{-6.15}^{+6.25}$ keV and $kT_{\rm max}=807.00_{-113.78}^{+123.29}$ keV, although the CPL model also gives an acceptable fit with photon index $\Gamma_{\rm ph}=-0.28_{-0.08}^{+0.06}$ and peak energy $E_{\rm p}=1118.09_{-75.49}^{+113.39}$ keV. The photon count spectra, modeled $f_{\rm \nu}$ spectra, and corner diagrams for the parameter constraints of the two fits are shown in Figure \ref{fig:mbb_corner} and Figure \ref{fig:cpl_corner}, respectively. For the weak tail in the time range from $T_0+0.005$ s to $T_0+0.20$ s, the CPL model provides the best fit with a relatively smaller peak energy, and the mBB model gives the second-best fit with a lower temperature. The total time-integrated spectrum measured between $T_0-0.005$ s and $T_0+0.20$ s can be described best by the mBB model with $kT_{\rm min}=34.29_{-4.79}^{+8.88}$ keV and $kT_{\rm max}=542.61_{-54.22}^{+71.79}$ keV, and the CPL model also gives an acceptable fit with photon index $\Gamma_{\rm ph}=-0.14_{-0.06}^{+0.06}$ and peak energy $E_{\rm p}=926.68_{-52.33}^{+51.78}$ keV. The BB model is significantly disfavored in the time-integrated spectra according to the BIC.

\begin{splitdeluxetable*}{clcllrccBcrllccrcc}
\tablecaption{Both time-integrated and time-dependent spectral fittings of GRB 200415A. \label{tab:spe_para}}
\tablehead{
\colhead{Time Intervals} & \colhead{Best} & \colhead{Flux} & \multicolumn{5}{c}{mBB Parameters} & \colhead{Time Intervals} & \multicolumn{4}{c}{CPL Parameters} & & \multicolumn{3}{c}{BB Parameters} \\
\cline{4-8}
\cline{10-13}
\cline{15-17}
\colhead{($t_1$, $t_2$) (s)} & \colhead{Model} & \colhead{(${\rm erg\ cm^{-2}\ s^{-1}}$)} & \colhead{$kT_{\rm min}$ (keV)} & \colhead{$kT_{\rm max}$ (keV)} & \colhead{$m$} & \colhead{pgstat/dof} & \colhead{BIC} & \colhead{($t_1$, $t_2$) (s)} & \colhead{$\Gamma_{\rm ph}$} & \colhead{$E_{\rm p}$ (keV)} & \colhead{pgstat/dof} & \colhead{BIC} & & \colhead{$kT$ (keV)} & \colhead{pgstat/dof} & \colhead{BIC}
} 
\startdata 
(-0.005, 0.005) & mBB & $4.32_{-0.62}^{+0.66}\times 10^{-4}$ & $39.40_{-6.15}^{+6.25}$ & $807.00_{-113.78}^{+123.29}$ & $-0.33_{-0.15}^{+0.17}$ & $278.5/350$ & $302.01$ & (-0.005, 0.005) & $-0.28_{-0.08}^{+0.06}$ & $1118.09_{-75.49}^{+113.39}$ & $300.2/351$ & $317.79$ & & $140.45_{-6.83}^{+8.48}$ & $458.4/352$ & $470.10$ \\
(0.005, 0.200) & CPL & $2.79_{-0.27}^{+0.32}\times 10^{-5}$ & $27.75_{-5.48}^{+15.31}$ & $399.87_{-43.70}^{+81.42}$ & $0.38_{-0.40}^{+0.17}$ & $277.8/350$ & $301.28$ & (0.005, 0.200) & $-0.01_{-0.08}^{+0.09}$ & $826.43_{-52.00}^{+59.65}$ & $279.1/351$ & $296.68$ & & $143.09_{-5.38}^{+5.27}$ & $385.3/352$ & $397.05$ \\
(-0.005, 0.200) & mBB & $4.53_{-0.44}^{+0.45}\times 10^{-5}$ & $34.29_{-4.79}^{+8.88}$ & $542.61_{-54.22}^{+71.79}$ & $-0.00_{-0.23}^{+0.15}$ & $279.9/350$ & $303.34$ & (-0.005, 0.200) & $-0.14_{-0.06}^{+0.06}$ & $926.68_{-52.33}^{+51.78}$ & $292.3/351$ & $309.92$ & & $142.12_{-4.03}^{+4.36}$ & $533.5/352$ & $545.27$ \\
\hline
(-0.005, -0.003) & CPL & $1.99_{-0.39}^{+0.50}\times 10^{-4}$ & $43.32_{-3.98}^{+11.64}$ & $402.21_{-102.98}^{+814.98}$ & $-1.18_{-0.58}^{+0.31}$ & $192.1/350$ & $215.55$ & (-0.005, -0.003) & $0.36_{-0.31}^{+0.29}$ & $393.53_{-38.06}^{+72.77}$ & $196.6/351$ & $214.18$ & & $82.27_{-6.53}^{+9.69}$ & $205.5/352$ & $217.19$ \\
(-0.003, -0.001) & mBB & $4.10_{-0.93}^{+1.58}\times 10^{-4}$ & $47.96_{-5.94}^{+13.75}$ & $539.31_{-84.08}^{+501.25}$ & $-0.85_{-0.57}^{+0.22}$ & $208.5/350$ & $231.98$ & (-0.003, -0.001) & $0.03_{-0.18}^{+0.22}$ & $607.43_{-75.55}^{+109.26}$ & $217.7/351$ & $235.26$ & & $97.81_{-6.84}^{+10.26}$ & $237.5/352$ & $249.27$ \\
(-0.001, 0.001) & CPL & $6.61_{-1.66}^{+2.19}\times 10^{-4}$ & $42.29_{-4.45}^{+44.52}$ & $789.59_{-118.46}^{+343.87}$ & $0.45_{-0.67}^{+0.20}$ & $221.1/350$ & $244.58$ & (-0.001, 0.001) & $-0.00_{-0.16}^{+0.26}$ & $1688.27_{-224.37}^{+304.76}$ & $222.6/351$ & $240.22$ & & $299.57_{-42.39}^{+0.40}$ & $241.2/352$ & $252.97$ \\
(0.001, 0.005) & BB & $1.79_{-0.38}^{+0.46}\times 10^{-4}$ & $94.05_{-29.75}^{+1.31}$ & $533.12_{-162.44}^{+257.78}$ & $-0.76_{-0.30}^{+1.25}$ & $206.2/350$ & $229.64$ & (0.001, 0.005) & $0.60_{-0.26}^{+0.46}$ & $857.35_{-132.59}^{+134.64}$ & $208.6/351$ & $226.25$ & & $182.53_{-18.18}^{+27.82}$ & $212.4/352$ & $224.10$ \\
(0.005, 0.010) & BB & $1.17_{-0.32}^{+0.37}\times 10^{-4}$ & $68.85_{-19.77}^{+26.53}$ & $298.35_{-60.56}^{+281.16}$ & $1.64_{-1.81}^{+0.00}$ & $177.1/350$ & $200.60$ & (0.005, 0.010) & $1.67_{-0.68}^{+0.88}$ & $847.03_{-121.32}^{+198.13}$ & $176.4/351$ & $194.05$ & & $233.18_{-29.92}^{+32.22}$ & $177.0/352$ & $188.73$ \\
(0.010, 0.020) & CPL & $1.16_{-0.20}^{+0.26}\times 10^{-4}$ & $77.42_{-18.98}^{+16.68}$ & $375.41_{-19.63}^{+225.17}$ & $0.63_{-1.22}^{+0.18}$ & $218.4/350$ & $241.91$ & (0.010, 0.020) & $0.53_{-0.20}^{+0.30}$ & $907.19_{-89.61}^{+99.54}$ & $220.4/351$ & $238.05$ & & $202.23_{-14.30}^{+21.71}$ & $227.9/352$ & $239.66$ \\
(0.020, 0.040) & BB & $5.07_{-0.75}^{+0.86}\times 10^{-5}$ & $98.56_{-38.76}^{+1.44}$ & $273.80_{-1.52}^{+198.59}$ & $0.81_{-1.68}^{+0.18}$ & $194.2/350$ & $217.66$ & (0.020, 0.040) & $0.63_{-0.20}^{+0.38}$ & $743.53_{-75.94}^{+74.67}$ & $194.6/351$ & $212.19$ & & $168.98_{-12.12}^{+15.06}$ & $198.8/352$ & $210.55$ \\
(0.040, 0.080) & CPL & $3.76_{-0.54}^{+0.66}\times 10^{-5}$ & $34.33_{-18.48}^{+43.53}$ & $269.64_{-1.92}^{+187.74}$ & $0.81_{-1.65}^{+0.43}$ & $257.4/350$ & $280.83$ & (0.040, 0.080) & $0.31_{-0.16}^{+0.23}$ & $676.90_{-58.29}^{+66.68}$ & $257.9/351$ & $275.50$ & & $139.89_{-7.78}^{+10.02}$ & $273.9/352$ & $285.62$ \\
(0.080, 0.120) & BB & $8.38_{-1.48}^{+1.80}\times 10^{-6}$ & $48.89_{-5.89}^{+16.36}$ & $221.49_{-61.68}^{+697.44}$ & $-0.83_{-0.96}^{+0.95}$ & $195.2/350$ & $218.67$ & (0.080, 0.120) & $0.65_{-0.35}^{+0.52}$ & $374.23_{-46.83}^{+63.24}$ & $196.4/351$ & $214.00$ & & $85.80_{-7.34}^{+10.20}$ & $199.0/352$ & $210.70$ \\
(0.120, 0.200) & PL & $3.78_{-2.48}^{+2.93}\times 10^{-6}$ & \multicolumn{5}{c}{Unconstrained} & (0.120, 0.200) & \multicolumn{4}{c}{Unconstrained} & & \multicolumn{3}{c}{Unconstrained} \\
\hline
\hline
\enddata
\tablecomments{The CPL model can be expressed as $N(E)=AE^{\Gamma_{\rm ph}}exp[-E(2+\Gamma_{\rm ph})/E_{\rm p}]$. The PL model gives an acceptable fit in the time slice between $T_0+0.12$ s and $T_0+0.20$ s: $\Gamma_{\rm ph}=-1.44_{-0.28}^{+0.11}$, pgstat/dof=179.2/352, BIC=190.96. Flux is derived based on the best model within 10--10,000 keV for each slice. Here the errors correspond to the 1$\sigma$ credible intervals.}
\end{splitdeluxetable*}

To study the spectral evolution, we perform the time-dependent spectral analysis for each of the ten slices mentioned above. The spectrum measured between $T_0-0.001$ s and $T_0+0.001$ s characterizing the hardest part of the burst can be fitted excellently by the CPL model with photon index $\Gamma_{\rm ph}=-0.00_{-0.16}^{+0.26}$ and peak energy $E_{\rm p}=1688.27_{-224.37}^{+304.76}$ keV. The mBB model gives a secondary fit with $kT_{\rm min}=42.29_{-4.45}^{+44.52}$ keV and $kT_{\rm max}=789.59_{-118.46}^{+343.87}$ keV. The best-fit models of the time-dependent spectra are generally the BB or CPL model, except for the second spectrum (mBB) and the last spectrum, which can be fitted by the PL model.

\begin{figure*}
 \label{fig:spectral_evolution}
 \centering
 \includegraphics[width=1\textwidth]{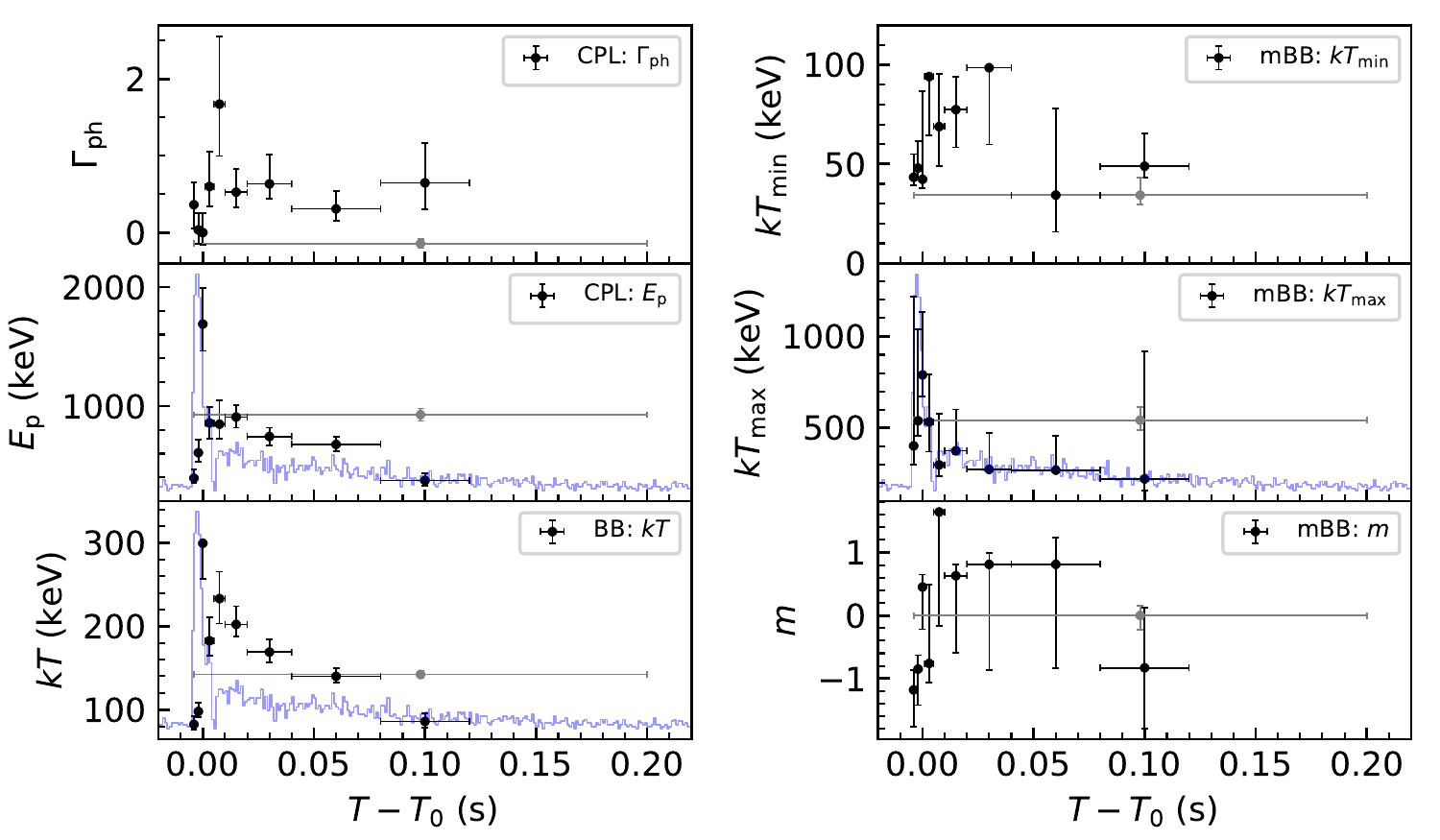} 
 \caption{Evolution of spectral parameters. The horizontal errors indicate the lengths of the time slices. The vertical errors correspond to the 1$\sigma$ uncertainties. For comparison, the light curve obtained by summing GBM detectors n1 and n3 is shown with a blue line in several panels. The spectral parameters of the final time interval fitted by these three models are excluded due to unconstrained fitting results.}
\end{figure*}

Following the first time-dependent spectrum, strong spectral evolution is observed until $T_{0}+0.2$ s where the spectra are smeared into the background. Figure \ref{fig:spectral_evolution} shows the evolution of the parameters of the CPL, BB, and mBB models. For comparison, the light curve obtained by summing GBM detectors n1 and n3 is overplotted in several panels. It is clear that the behaviors of $E_{\rm p}$ and the temperature of this burst show an intensity tracking pattern \citep{Golenetskii:1983Natur.306.451G}, even though there is a delay of $\sim2$ ms between the peaks of the flux and the spectral evolution. During the first spike, the rapid evolution of the peak energy and temperature imply that the emission source underwent an abrupt variation. Subsequent decay in the weak tail stage may indicate the cooling of the emission region. Such spectral evolution is consistent with a rapidly expanding then gradually cooling fireball-like emission source.

\subsection{Burst Energy}

Based on the spectral analyses, the average flux is derived from the best model within 10--10,000 keV for each slice, as shown in Table \ref{tab:spe_para}. The total fluence in the time range from $T_0-0.005$ s to $T_0+0.20$ s is $9.29_{-0.90}^{+0.92}\times 10^{-6}\ \rm {erg\ cm^{-2}}$. The peak flux of $1.11_{-0.11}^{+0.15}\times 10^{-3}\ \rm {erg\ cm^{-2}\ s^{-1}}$ is approximately proportional to average flux $F_{\rm ave}$ with a scale factor $R_{\rm peak}/R_{\rm ave}$, where $R_{\rm peak}$ is 0.4-ms peak count rate. Assuming a distance of 3.5 Mpc, the corresponding isotropic energy and peak luminosity are estimated as $E_{\gamma,\rm iso}=1.36_{-0.13}^{+0.14}\times 10^{46}$ erg and $L_{\gamma,\rm p,iso}=1.62_{-0.16}^{+0.21}\times 10^{48}$ $\rm erg\ s^{-1}$, respectively. The above values, along with some important temporal and spectral observational properties, are summarized in Table \ref{tab:para}.

\subsection{Spectral Hardness}

\begin{figure}
 \label{fig:GRB200415A_HR}
 \centering
 \includegraphics[width=0.47\textwidth]{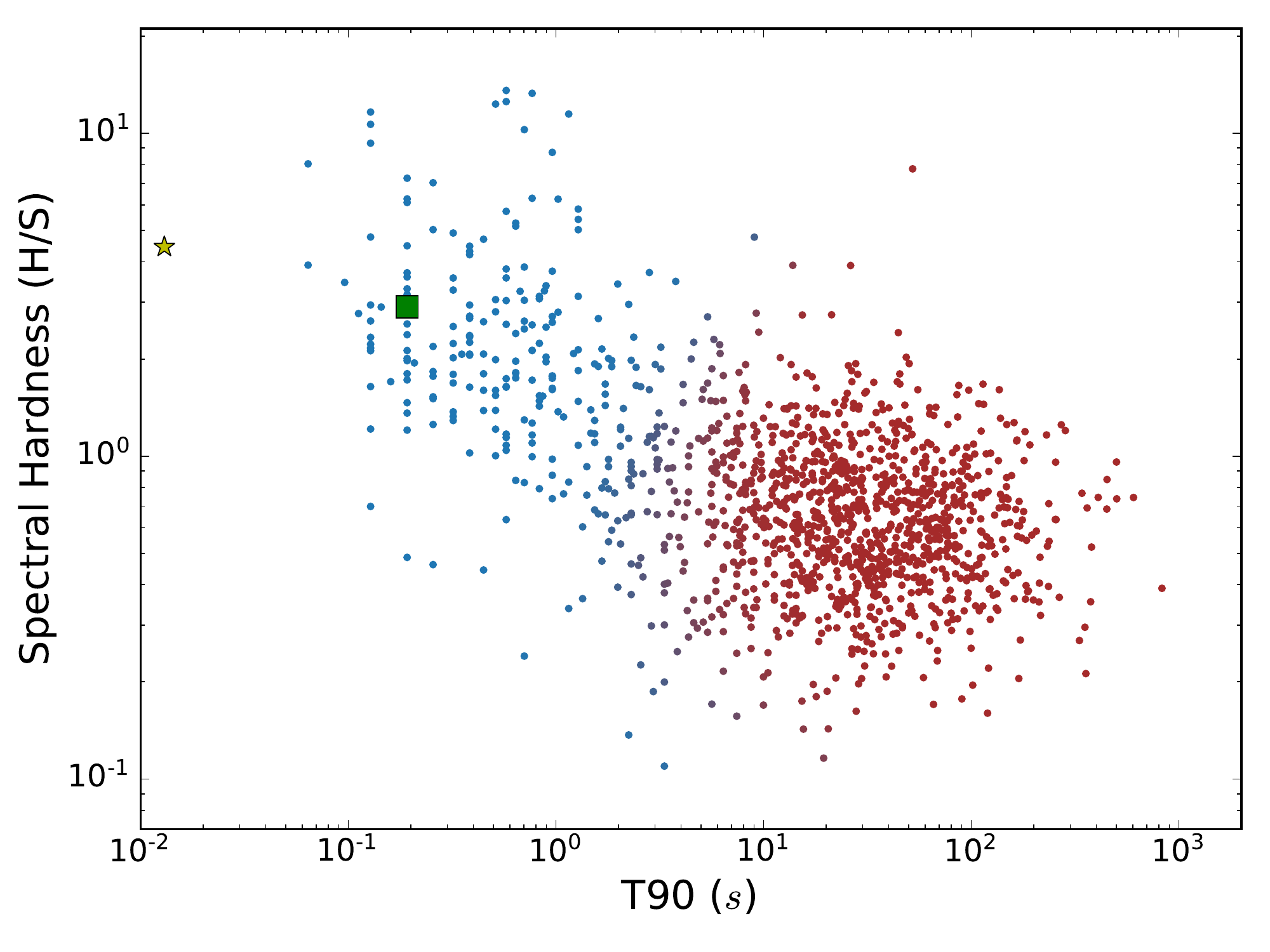}
 \caption{The $T_{90}$--HR diagram. The sample of short (blue circles) and long (red circles) GRBs is obtained from \cite{Goldstein:2017ApJ}. First spike (star) of GRB 200415A: HR = 4.45; soft emission (square) of GRB 200415A: HR = 2.3. Here HR = H/S = (50--300 keV)/(10--50 keV). The first spike is considered from $T_0-0.005$ s to $T_0+0.008$ s, and soft emission is considered from $T_0+0.008$ s to $T_0+0.20$ s.}
\end{figure}

Short GRBs have more photons at higher energy bands \citep{Dezalay:1991AIPC}. This can be quantitatively represented by the ratio of observed counts in 50--300 keV (H) to 10--50 keV (S). We compute the spectral hardness (H/S) for the observed spike and tail individually. The spike is considered from $T_0-0.005$ s to $T_0+0.008$ s, and the subsequent tail is considered from $T_0+0.008$ s to $T_0+0.20$ s. The results are plotted in Figure \ref{fig:GRB200415A_HR} with a sample of short (blue circles) and long (red circles) GRBs obtained from \cite{Goldstein:2017ApJ}. Both the spike and the weak tail are slightly harder than most short GRBs with H/S $\sim$ 4.45 and 2.3, respectively.

\subsection{Fermi-LAT Detection}\label{sec:this_section}

We extracted the Fermi-LAT data within a temporal window extending 1000 s after $T_0$. Then we performed an unbinned likelihood analysis. The data were filtered by selecting photons with energies in the range 100 MeV--300 GeV and within a region of interest (ROI) of $12^{\circ}$ radius centered on the burst position. A further selection of zenith angle ($\rm 100^{\circ} $) was applied to reduce the contamination of photons coming from the Earth limb. We adopt the P8R3\_TRANSIENT020E\_V2 response, which is suitable for faint detection. The probabilities of the photons being associated with the source are calculated using the \sw{gtsrcprob} tool. The highest-energy photon is a 1.72 GeV event, which is observed 284 s after the GBM trigger. All LAT photon events are plotted in Figure \ref{fig:bb_lcs}, and the most significant five LAT photon events are listed in Table \ref{tab:lat_sed}. For the time-integrated duration of 0--1000 s, the energy and photon flux in the 0.1--10 GeV energy range are $(3.78\pm2.24)\times 10^{-9}$ $\rm erg$ $\rm s^{-1}$ $\rm cm^{-2}$ and $(3\pm1.74)\times 10^{-6}$ $\rm photons$ $\rm s^{-1}$ $\rm cm^{-2}$, respectively. The spectral index is $-1.47 \pm 0.43$ with a test statistic of detection of 26.

\section{Physical Origin: A Typical Short GRB or an SGR GF?} \label{sec:physical_origin}

In this section, we discuss various observational evidence and physical conditions that can support GRB 200415A as either a typical short GRB or a GF from a magnetar.

\subsection{Placement of the Event}

\begin{figure}
 \label{fig:ep_eiso}
 \centering
 \includegraphics[width=0.45\textwidth]{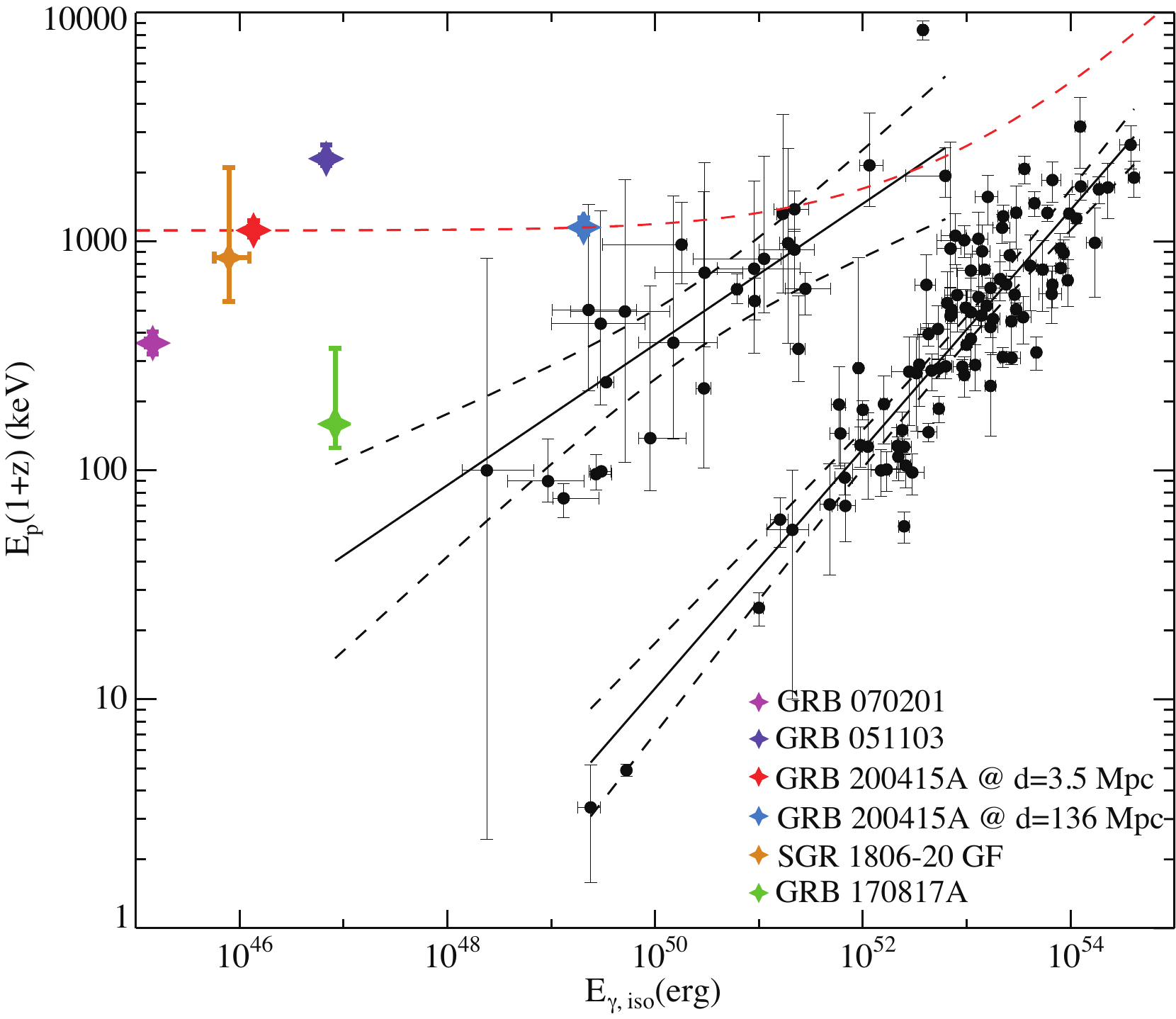}
 \caption{The $E_{\rm p}$ and $E_{\rm\gamma,iso}$ correlation diagram. The upper and lower black solid lines show the best-fit correlations for the short and long GRB population, respectively. The GRB 200415A with different redshifts follows the track shown with the red dashed line. The SGR 1806-20 GF, GRB 170817A, and two GF candidates (GRB 051103 and GRB 070201) are also plotted here.}
\end{figure}

Assuming the association with the Sculptor galaxy is real, we can directly check if GRB 200415A exhibits differently when compared with other GRBs. To do so, we place GRB 200415A in the $E_{\rm p}$--$E_{\rm\gamma,iso}$ diagram (aka the Amati relation), in which long and short GRBs typically follow different tracks. In Figure \ref{fig:ep_eiso}, we overplot the Amati relation using the GRBs with known redshift \citep{Amati2002, Zhang:2009ApJ.703.1696Z}. Short GRBs follow the track of log$E_{\rm p,z}$ = $a$ + $b$log$E_{\rm\gamma,iso}$ with the best-fitting parameters being $a=-12.82$ and $b=0.31$, where $E_{\rm p,z}=E_{\rm p}(1+z)$, while GRB 200415A, with different redshifts, follows the track shown with the red dashed line. Assuming a distance of 3.5 Mpc, it is clear that GRB 200415A significantly deviates from such a correlation. To be on the short GRB track, GRB 200415A would have to be much further, with $z\gtrsim0.03$ ($d\gtrsim136$ Mpc). Moreover, compared with the off-axis GRB 170817A, which is a nearby event at $\sim40$ Mpc, GRB 200415A obviously possesses a lower energy and a much higher $E_{\rm p}$. On the other hand, by plugging in the properties of the giant outburst from SGR 1806-20, we find that GRB 200415A is very close to its location, which points toward the GF origin.

\begin{table}
\begin{center}
\caption{Observational properties of GRB 200415A. The total fluence and peak flux are calculated in 10--10,000 keV energy band. Here the time-integrated spectral parameters are measured over the total duration. All errors correspond to the 1$\sigma$ credible intervals.}
\label{tab:para}
\begin{tabular}{ll}
\hline
\hline
Observed Properties & GRB 200415A \\
\hline
Abrupt rise time & $\sim 2$ ms \\
Steep decay time & $\sim 8$ ms \\
$T_{\rm 90}$ (sharp peak only) & $5.88_{-0.34}^{+0.23}$ ms \\
Total duration & $\sim200$ ms \\
$\Gamma_{\rm ph}$ at peak & $-0.00_{-0.16}^{+0.26}$ \\
$E_{\rm p}$ at peak & $1688.27_{-224.37}^{+304.76}$ keV \\
Time-integrated $\Gamma_{\rm ph}$ & $-0.14_{-0.06}^{+0.06}$ \\
Time-integrated $E_{\rm p}$ & $926.68_{-52.33}^{+51.78}$ keV \\
Total fluence & $9.29_{-0.90}^{+0.92}\times10^{-6}$ $\rm erg\ cm^{-2}$ \\
Peak flux & $1.11_{-0.11}^{+0.15}\times 10^{-3}\ \rm {erg\ cm^{-2}\ s^{-1}}$ \\
Possible host galaxy & Sculptor galaxy (NGC 253) \\
Distance & 3.5 Mpc \\
Isotropic energy $E_{\rm\gamma,iso}$ & $1.36_{-0.13}^{+0.14}\times 10^{46}$ erg \\
Peak luminosity $L_{\rm\gamma,p,iso}$ & $1.62_{-0.16}^{+0.21}\times 10^{48}$ erg s$^{-1}$ \\
\hline
\hline
\end{tabular}
\end{center}
\end{table}

\subsection{Can We Observe It as GRB 200415A if There Is a Compact Star Merger Event at d = 3.5 Mpc?}

It is natural to ask a question: if a compact star merger event, as an origin of typical short GRBs, happens at $d$ = 3.5 Mpc, what would be the observational consequence, and how similar is it compared with GRB 200415A? Straightforwardly, considering a typical short GRB with $E_{\rm\gamma,iso}=10^{49}$--$10^{52}$ erg \citep{Zhang:2009ApJ.703.1696Z, Gruber_2014}, the corresponding fluence at $d$ = 3.5 Mpc is $\sim10^{-3}$--$10^{0}$ erg cm$^{-2}$, which is at least three orders of magnitude higher than the observed one in GRB 200415A. This suggests that GRB 200415A has to be an off-axis GRB so that the observed fluence can be significantly reduced due to the beaming effect if it is indeed from a compact star merger.

However, is an off-axis merger-type short GRB consistent with the observations of GRB 200415A? The isotropic equivalent energy of GRB 200415A is $\sim 10^{46}$ erg, the $E_{\rm p}$ is $\sim1$ MeV, and its sharp peak duration is $\sim 6$ ms. If it is a typical short burst seen off-axis, assuming the top-hat model, the relationship between duration (or peak energy) and the Doppler factor is approximately \citep{Granot:2002ApJ.570L.61G, Abbott:2017ApJ.848L.13A}
\begin{equation}
 \frac{T_{90,\rm{off}}}{T_{90,\rm{on}}}=\frac{E_{\rm{p,on}}}{E_{\rm{p,off}}}=\frac{D(0)}{D(\theta_{\rm v}-\theta_{\rm j})}=a\approx 1+\Gamma^2(\theta_{\rm v}-\theta_{\rm j})^2, 
\end{equation}
where $\Gamma$ is the Lorentz factor of the jet, $D$ is the Doppler factor, and $\theta_{\rm j}$ and $\theta_{\rm v}$ are the jet opening angle and viewing angle, respectively. Note that the ratio of duration applies to the case of a single pulse \citep{Zhang:2009ApJ.703.1696Z} or one central engine activity, and this burst is the case. For multipulse bursts, if the duration of one pulse is much larger than the duration of the central engine activity, this ratio also approximately applies. The isotropic equivalent energy scales as $\propto a^{-2}$ for $\theta_{\rm j}<\theta_{\rm v}<2\theta_{\rm j}$ and $\propto a^{-3}$ for $\theta_{\rm v}>2\theta_{\rm j}$ \citep{Ioka:2018PTEP}. Assuming that the on-axis isotropic energy is the typical value of $E_{\rm{\gamma,iso}}=10^{50}$ erg for short bursts, we have $a\sim100$ for $\theta_{\rm j}<\theta_{\rm v}<2\theta_{\rm j}$ or $a\sim20$ for $\theta_{\rm v}>2\theta_{\rm j}$. So, the on-axis duration and peak energy would be $T_{90}=0.06(a/100)^{-1}$ ms or $0.3(a/20)^{-1}$ ms and $E_{\rm p}=100(a/100)$ MeV or $20(a/20)$ MeV. These values are obviously atypical and have never been observed in existing short GRB samples.

Another possibility is that GRB 200415A is off-axis from a structured jet that includes a uniformly bright core surrounded by a PL or Gaussian decaying wing \citep{Rossi:2002MNRAS.332.945R, Zhang:2002ApJ.571.876Z}. In this model, the energy/luminosity, Lorentz factor, and $E_{\rm p}$ all depend on the angle from the jet axis. The high $E_{\rm p}$ ($\sim$ 1 MeV) and unusually low $E_{\rm\gamma,iso}$ ($\sim 10^{46}$ erg) observed in GRB 200415A require that the energy profile is quite steep while the $E_{\rm p}$ profile is flat. Such conditions are at odds with most studies of GRB jets, although they can not be fully ruled out.

\begin{table}
\begin{small}
\caption{LAT-HE observation in 0--1000 s. The time of arrival of the photons since the GBM trigger time, the energy and the probability of association $>80\%$ with the source are listed.}
\label{tab:lat_sed}
\begin{center}
\begin{tabular}{ccc}
\hline
Time & Energy & Probability\\
(s) & (GeV) & (\%) \\ \hline
19.18 & 0.48 & 99 \\ 
180.22 & 1.32 & 99 \\
276.88 & 0.53 & 87\\
284.06 & 1.72 & $\sim100$\\
471.16 & 0.14 & 90\\ \hline
\end{tabular}
\end{center}
\end{small}
\end{table}

A similar case to GRB 200415A is GRB 170817A when considering the nearby distance and the explanation using a structured jet model \citep[e.g.,][]{Meng:2018ApJ.860.72M}, which was associated with the gravitational-wave event GW170817 that originated from a compact star merger. It is interesting to compare the two events. The isotropic energy of GRB 200415A is similar to that of GRB 170817A (also $\sim 10^{46}$ erg). Therefore, the jet kinetic energies should be analogical for both events if the radiation efficiency is the same. We can further speculate that the progenitors of the two bursts are also similar. The GRB 170817A was followed by a ``kilonova" that was observed about half a day after the burst \citep{Coulter:2017Sci.358.1556C} and lasted for tens of days, while to date, no follow-up emission in any bands has been reported for GRB 200415A, which is one order of magnitude closer than GRB 170817A. Note that the emission of kilonovae is approximately isotropic, so different viewing angles do not affect the detection of kilonovae. If GRB 200415A was from a structured jet seen off-axis, its afterglow with a long time rise should also have been observed, like GRB 170817A, due to its similar jet energy and nearer distance. Thus, the nondetection of the kilonova and afterglow of GRB 200415A does not favor it as a typical short GRB.

In summary, if the association of the host galaxy is real, it is unlikely that GRB 200415A was generated from a compact star merger event.

\subsection{Is GRB 200415A an SGR GF?}

For quite some time, GFs from SGRs in nearby galaxies have been proposed to serve a few misclassified short GRBs since a long time ago \citep{Ofek:2006ApJ.652.507O, Ofek:2008ApJ, Frederiks:2007AstL, Mazets:2008ApJ}. The claims are mainly based on the consistency with the energies of GFs and the associations with some nearby galaxies \citep[e.g., M31;][]{Ofek:2008ApJ}. In fact, the total energy ($\sim$ 10$^{46}$ erg) of GRB 200415A and its location near the Sculptor galaxy are in agreement with such a scenario. In this section, we perform additional statistical studies and discuss some theoretical evidence to support the GF origin of GRB 200415A.

\begin{table*}
\begin{small}
\begin{center}
\caption{Comparison among GRB 200415A, the SGR 1806-20 GF, GRB 051103, and GRB 070201. The spectral properties of GRB 200415A are given by the best models of time-integrated spectra. The distance of $\sim8.7$ kpc to the location of SGR 1806-20 is suggested by \cite{Bibby:2008MNRAS.386L.23B}.}
\label{tab:gf_vs_200415}
\begin{tabular}{ccccc}
\hline
\hline
Properties & GRB 200415A & SGR 1806-20 GF & GRB 051103 & GRB 070201 \\
\hline
Location & NGC 253 & Massive star cluster & M81 & M31 \\
Distance & 3.5 Mpc & 8.7 kpc & 3.6 Mpc & 0.78 Mpc \\
\hline
\textbf{Initial Pulse} & & & & \\
Steep rise, ms & $\sim 2$ & $\sim 1$ & $\le 6$ & $\sim 20$ \\
Decay, ms & $\sim 8$ & $\sim 200$ & $\sim 40$ & $\sim 160$ \\
Rapid spectral evolution & \ding{51} & \ding{51} & \ding{51} & \ding{51} \\
CPL photon index $\Gamma_{\rm ph}$ & $-0.28_{-0.08}^{+0.06}$ & $-0.73_{-0.47}^{+0.64}$ & $0.16_{-0.15}^{+0.19}$ & $-0.52_{-0.13}^{+0.15}$ \\
CPL peak energy, keV & $1118.09_{-75.49}^{+113.39}$ & $850_{-303}^{+1259}$ & $2300_{-150}^{+350}$ & $360_{-38}^{+44}$ \\
BB temperature, keV & $140.45_{-6.83}^{+8.48}$ & $175\pm 25$, 116 & ... & ... \\
Peak flux, $\rm {erg\ cm^{-2}\ s^{-1}}$ & $1.11_{-0.11}^{+0.15}\times 10^{-3}$ & $\sim 5.0$, $13.1_{-4.4}^{+8.0}$ & $(2.8\pm 0.3)\times 10^{-3}$ & $1.61_{-0.50}^{+0.29} \times 10^{-3}$ \\
Peak luminosity, $\rm erg\ s^{-1}$ & $1.62_{-0.16}^{+0.21}\times 10^{48}$ & $0.45 (1.19)\times 10^{47}$ & $(4.34\pm 0.46)\times 10^{48}$ & $1.2\times 10^{47}$ \\
\hline
\textbf{Tail} & & & & \\
Tail duration & $\sim 150$ ms & $\sim 380$ s & $\sim 130$ ms & $\sim 100$ ms \\
Period, s & ... & 7.56 & ... & ... \\
QPO & ... & \ding{51} & ... & ... \\
BB temperature, keV & $143.09_{-5.38}^{+5.27}$ & $\sim 30$ & ... & ... \\
CPL photon index $\Gamma_{\rm ph}$ & $-0.01_{-0.08}^{+0.09}$ & ... & $0.43_{-0.40}^{+0.34}$ & $\sim(-1)$ \\
CPL peak energy, keV & $826.43_{-52.00}^{+59.65}$ & ... & $530\pm 80$ & $\sim 125$ \\
Fluence, $\rm erg\ cm^{-2}$ & $5.44_{-0.52}^{+0.63}\times 10^{-6}$ & $8\times 10^{-3}$ & $(2\pm0.3)\times 10^{-6}$ & $\sim10^{-6}$ \\
\hline
Total fluence, $\rm erg\ cm^{-2}$ & $9.29_{-0.90}^{+0.92}\times10^{-6}$ & $0.87_{-0.24}^{+0.50}$ & $(4.4\pm 0.5)\times 10^{-5}$ & $2_{-0.26}^{+0.10}\times 10^{-5}$ \\ 
Total energy, erg & $1.36_{-0.13}^{+0.14}\times 10^{46}$ & $7.88_{-2.17}^{+4.53}\times 10^{45}$ & $(6.82\pm0.78)\times 10^{46}$ & $1.5_{-0.19}^{+0.07}\times 10^{45}$ \\
\hline
References & & \cite{Hurley:2005Nature} & \cite{Frederiks:2007AstL33} & \cite{Mazets:2008ApJ} \\
 & & \cite{Frederiks:2007AstL} & \cite{Ofek:2006ApJ.652.507O} & \cite{Ofek:2008ApJ} \\
\hline
\hline
\end{tabular}
\end{center}
\end{small}
\end{table*}

To compare with previous GF-sGRB candidates and known SGR GF events, we list some detailed properties of the GF-sGRBs 200415A, 051103, and 070201, as well as the SGR 1806-20 GF in Table \ref{tab:gf_vs_200415}. They are all located near (or in) galaxies or massive star clusters with high star formation rates. In terms of isotropic energy and peak luminosity, they are also analogous when we assume the distance of GRB 200415A is 3.5 Mpc. Interestingly, as we show in Figure \ref{fig:ep_eiso}, GRB 200415A at a luminosity distance of 3.5 Mpc seems to belong to the same population as the SGR 1806-20 GF and is closer to it than the GF-sGRBs 051103 and 070201.

\begin{figure}[t]
 \label{fig:grb_spec_para_dis}
 \includegraphics[width=0.45\textwidth]{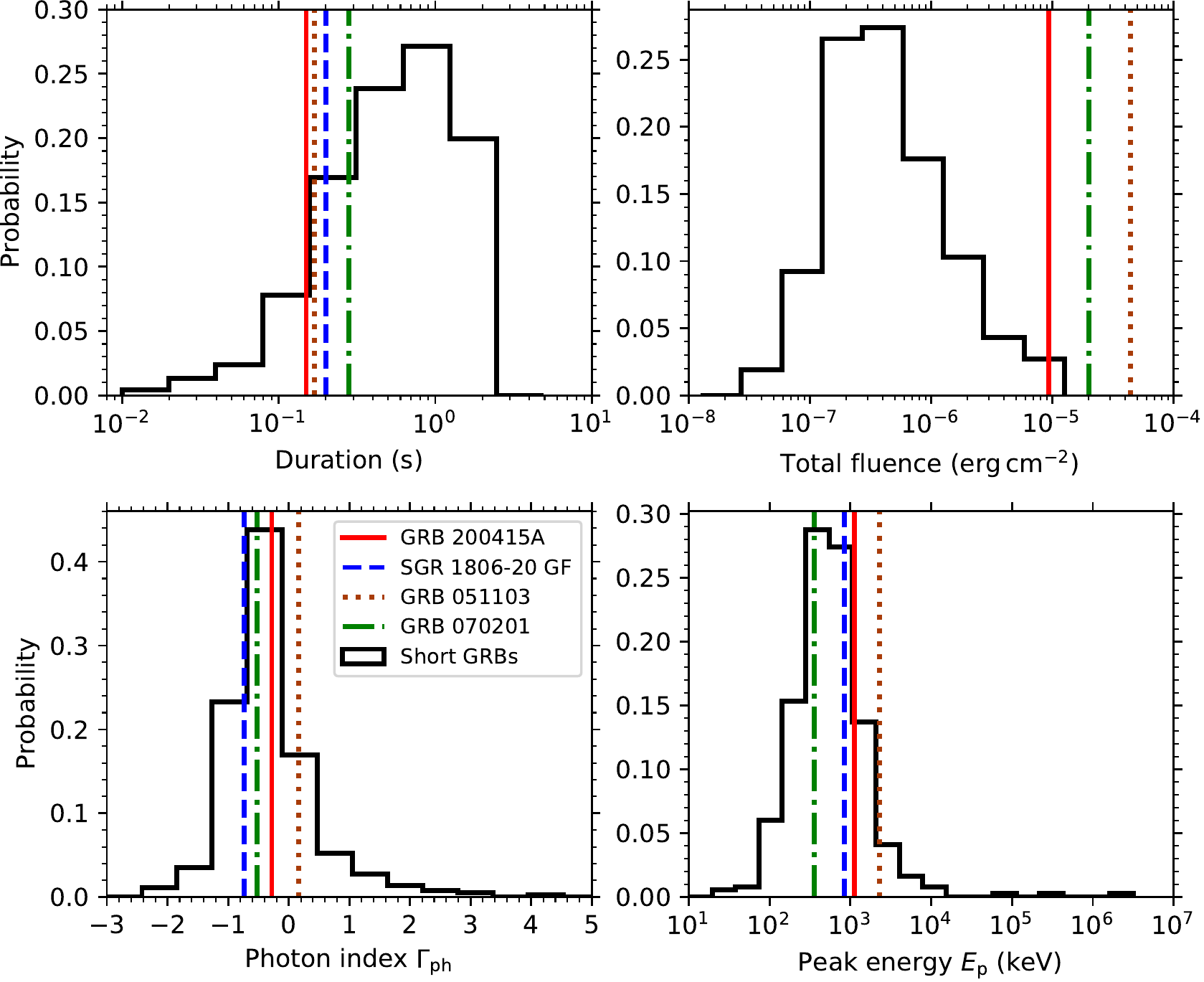}
 \caption{Probability distributions of the parameters of short GRBs. This short GRB sample is from the Fermi/GBM burst catalog. Here GRB 200415A, the GF from SGR 1806-20, GRB 051103, and GRB 070201 are placed in parameter space. The top left panel utilizes the total durations for GRB 200415A (150 ms), GRB 051103 (170 ms), and GRB 070201 (280 ms) but the duration of the initial pulse for the SGR 1806-20 GF (200 ms).}
\end{figure}

To further check if these GF-sGRBs are a special group in the whole short GRB sample, we overplot them in various distributions of characteristic parameters of short GRBs from the Fermi/GBM burst catalog \citep{Gruber_2014, von_Kienlin_2014, Bhat:2016ApJS, von_Kienlin_2020}. The GF from SGR 1806-20 is also marked, where applicable. As shown in Figure \ref{fig:grb_spec_para_dis}, the four events are clustered together in the distributions, with smaller $T_{\rm 90}$, significantly larger fluence, relatively typical photon index, and slightly higher $E_{\rm p}$ in the whole short GRB sample. In addition, the four events are also clustered in the $E_{\rm p}$--$E_{\rm iso}$ diagram (Figure \ref{fig:ep_eiso}). All of the above statistical similarities suggest that GRB 200415A should originate from an SGR GF.

Theoretically speaking, generating an sGRB-like event is plausible in the context of the GF model. Large-scale shearing and reconnection of an external magnetic field stronger than $\sim 10^{14}$ G \citep{Duncan:1992ApJ.392L}, which may be induced by internal instabilities, launch an expanding pair fireball \citep{Thompson:1995MNRAS}. The internal stored magnetic energy is released rapidly during the crossing time of the Alfven wave in the neutron star interior. The timescale that is consistent with the duration of the initial spike is $T_{\rm spike}\gtrsim R_{\rm NS}/V_{\rm A}\gtrsim 0.1\rm\,s$ with $R_{\rm NS}=10\rm\,km$, where $V_{\rm A}\lesssim B_{\rm p}/(4\pi\rho)^{1/2}$ is the Alfven velocity with a poloidal magnetic field $B_{\rm p}=10^{15}\rm\,G$ and density $\rho=10^{15}\rm\,g\,cm^{-3}$ \citep{Feroci:2001ApJ.549.1021F}. In another case, where the instabilities are driven by an external twisted magnetic field, the large-scale magnetic reconnection event will last $\sim10^2R_{\rm NS}/c\sim10^{-2}$ s \citep{Parfrey_2013}. Actually, the ``tail emission" of GRB 200415A is hard (see Table \ref{tab:gf_vs_200415}) and still presents a strong spectral evolution of hard to soft (see Figure \ref{fig:spectral_evolution}), like the decaying stages of the initial pulses of other GFs. As the previous observations suggested, the insignificant quasi-thermal spectral evolution was predicted in the soft tail of the GF \citep{Thompson:1995MNRAS}. It implies that the main spike and weak emission of GRB 200415A may be combined into the initial pulse of a typical GF. The energy released through the fireball should be $\lesssim 10^{47}B_{15}^2(R_{\rm NS}/10\rm\ km)^3$ erg, which is carried by an internal dipole field with $B=10^{15}$ G \citep{Thompson_2001}.

In the context of the GF model, the emissions could be quasi-thermal, as the mBB model fits the spectra of GRB 200415A well. The initial bright pulse is emitted by a sharply expanding pair fireball \citep{Thompson:1995MNRAS}. The radius of the neutron star gives a lower limit to the size of the initial fireball ($\gtrsim R_{\rm NS}=10$ km), which emits the maximum luminosity ($1.62\times10^{48}$ $\rm erg\,s^{-1}$). Then, the minimum time variability of 1.9 ms provides an upper limit to the size of the emission region ($\lesssim\Delta_{\rm t,min}c=570$ km) with an average luminosity of $\sim 10^{47}$ $\rm erg\,s^{-1}$. Assuming BB emission, we can derive the temperature range from $kT_{\rm max}\sim600$ keV to $kT_{\rm min}\sim40$ keV of the expanding fireball, which overlaps into an mBB spectrum. We note that such restrictions on temperature are similar to those of our mBB spectral fitting during the first spike (Table \ref{tab:spe_para}). On the other hand, for the peak time-dependent spectrum measured from $T_0-0.001$ s to $T_0+0.001$ s, the BB model can give an acceptable fit with $kT=299.57_{-42.39}^{+0.40}$ keV and flux $5.49_{-1.12}^{+0.80}\times10^{-4}$ $\rm erg\ cm^{-2}\ s^{-1}$. The corresponding radius of the emission source should be $27.80_{-2.84}^{+8.12}$ km, which is similar to the radii of the emission regions inferred in previous studies \citep[e.g.,][]{Nakar:2005ApJ.635.516N, Ofek:2006ApJ.652.507O, Ofek:2008ApJ}.

Assuming that GRB 200415A is an SGR GF at a distance of 3.5 Mpc, it is the first detection of GeV emission that is coincident with a GF. The first high-energy photon was detected at $\sim 19$ s after the onset of GRB 200415A, whose energy is $480$ MeV. The photon number spectrum during 0--1000 s in the 0.1--10 GeV band is $N(E)\sim1.4\times10^{-5}E^{-1.47}$ MeV$^{-1}$ s$^{-1}$ cm$^{-2}$ with a spectral index of $\sim -1.47$. The photon number that can annihilate the photons with an energy of $E_{\rm{max}}$ is $N_{>E_{\rm{max,an}}}=4\pi d_{\rm L}^2T\int_{E_{\rm{max,an}}}^{\infty}N(E)dE$, where $E_{\rm{max,an}}\equiv(\Gamma m_e c^2)^2/E_{\rm{max}}$ and $\Gamma$ is the Lorentz factor of the emission region. Such a photon number corresponds to the opacity for the $E_{\rm max}$ photons being $\tau\approx 0.1\sigma_T N_{>E_{\rm max,an}}/(4\pi R^2)$ \citep{Lithwick:2001ApJ.555.540L}. The emission region optically thin to the $E_{\rm max}$ photons requires that its radius should be $R\gtrsim3.8\times 10^{11}(E_{\rm max}/480{\rm\,MeV})^{0.235}(T/19\rm\,s)^{1/2}\Gamma^{-0.47}$ cm. The speed of the GeV emission region is unknown. If it is relativistic, taking $\Gamma=10$ for example, then $R\sim \Gamma^2 cT\approx6\times 10^{13}$ cm. This suggests that the delayed GeV emission region is much larger than the GF emission region, implying that there is an additional mechanism that is responsible for such GeV photons after the GF. A natural source is the afterglow-like emission from the outflow launched from the GF. For a relativistic outflow, the synchrotron or synchrotron self-Compton radiation from electrons generated from the forward shock can produce such a detectable sub-GeV or GeV afterglow emission \citep{Fan:2005MNRAS.361.965F}.

The energy of the GF was used to constrain the magnetic field of the magnetar \citep{Hurley:2005Nature}. In our scenario, we estimate an upper limit on $B_{\rm NS}$ using the main flare energy ($\sim 10^{46}$ erg). The magnetic field, $B$, within a radius $\Delta R \sim 10$ km, which is above the stellar radius $R_{\rm NS}$, is $B(R_{\rm NS} + \Delta R) \lesssim (8 \pi E_{\gamma, \rm{iso}}/3 \Delta R^3)^{1/2}$ \citep{Thompson:1995MNRAS}. At a radius $r$, the magnetic field, approximated as a dipole field, is $B(r) = B_{\rm NS} (r/R_{\rm NS})^{-3}$. Then, we constrain $B_{\rm NS} \lesssim 2\times 10^{15}$ G.

\section{Summary and Implication} \label{sec:summary}

Our analyses indicate that a giant flare (GF) from a soft gamma-ray repeater (SGR) provides the most natural explanation for the short gamma-ray burst (sGRB) GRB 200415A through the following facts.

\begin{enumerate}
 \item Good localization close to the Sculptor galaxy at $d$ = 3.5 Mpc.
 
 \item Similar energy to known GFs (e.g., from SGR 1806-20).
 
 \item A weak tail after the main spike.
 
 \item Tiny-lag result consistent with a one-time fireball emission region.
 
 \item Quasi-thermal spectra with a significant intensity tracking pattern of spectral evolution.
 
 \item Small duration, minimal time variability, energy, radius of emission region, and spectral temperature consistent with the magnetar model of a GF. 
 
 \item Significant outlier (lower energy and higher $E_{\rm p}$) away from the short GRB track in the $E_{\rm p}$--$E_{\rm iso}$ diagram but clustered together with other GF-sGRB candidates (GRB 051103 and GRB 070201) and the GF event from SGR 1806-20.
 
 \item Inconsistent with other short GRBs and contrived jet conditions if explained by a nearby compact star merger event that happened at $d=3.5$ Mpc. 
 
 \item Statistically similar to other GF-sGRB candidates and the observed GF from SGR 1806-20. 
\end{enumerate}

Nevertheless, our conclusion is subject to the coincidence of the association between GRB 200415A and the Sculptor galaxy. The chance possibility that they are not associated, unfortunately, can not be ruled out without a counterpart detected in other wavelengths. Future multiwavelength follow-up observations of this event are thus encouraged to confirm the SGR nature of this event.

The indistinguishableness of the spectral properties between GF-sGRB candidates and the whole sGRB sample (Figure \ref{fig:grb_spec_para_dis}) suggests that there might be some ultrashort GRBs that are actually GF-sGRBs at a large distance \citep{Tanvir:2005Natur}. Without host-galaxy measurements, such events are hard to identify, but the ultrashort duration may be a clue. A preliminary search of such ultrashort GRBs with $T_{\rm 90} < 0.1$ s has been done. Further more investigations of these events and follow-up observations of similar ones are essential to firmly establish the GF-sGRB connection.

\acknowledgments

B.B.Z acknowledges support by the Fundamental Research Funds for the Central Universities (14380035). This work is supported by the National Key Research and Development Programs of China (2018YFA0404204), the National Natural Science Foundation of China (grant Nos. 11833003, U1838105, and U1831135), and the Program for Innovative Talents, Entrepreneur in Jiangsu, and the Strategic Priority Research Program on Space Science, the Chinese Academy of Sciences, grant No. XDB23040400.

\clearpage

\end{document}